\documentclass{article}

\usepackage{amsthm}
\usepackage{amssymb}
\usepackage{amsmath}
\usepackage{amsfonts}

\numberwithin{equation}{section}

\begin{document}

\title{Coulomb integrals for the $SL(2,\mathbb{R})$ WZNW model}

\author{Sergio~M.~Iguri \footnote{e-mail: siguri@iafe.uba.ar} \, and
  Carmen~A.~N\'u\~nez
\footnote{e-mail: carmen@iafe.uba.ar}}

\date{\small Instituto de Astronom\'{\i}a y F\'{\i}sica del Espacio
  (CONICET-UBA).\\ 
C.~C.~67 - Suc.~28, 1428 Buenos Aires, Argentina \\ and \\
  Departamento de F\'\i sica, 
FCEN, Universidad de Buenos Aires. \\ Ciudad Universitaria, Pab. I,
  1428 Buenos Aires, 
Argentina.}

\maketitle

\begin{abstract}
We review the Coulomb gas computation of three-point functions in the
$SL(2,\mathbb{R})$ WZNW model and obtain explicit expressions for
generic states. These amplitudes have been computed in the past by
this and other methods but the analytic continuation in the number of
screening charges required by the Coulomb gas formalism  had only been performed
in particular cases. After showing that ghost contributions to the 
correlators can be
generally expressed in terms of Schur polynomials, we solve Aomoto integrals
in the complex plane, a new set of multiple integrals  of
Dotsenko-Fateev type. We then make use of monodromy  invariance to
analytically continue the number of screening operators and prove that
this procedure gives results in complete agreement with the amplitudes
obtained from the bootstrap approach. We also compute a four-point function
involving a spectral flow operator and we verify that it leads to  the one unit spectral
flow three-point function according to a prescription previously proposed
in the
literature. In addition,
we present an alternative method
to obtain spectral flow non-conserving $n-$point functions through
well defined operators and we prove that it reproduces the exact
correlators for $n=3$. Independence of the result on the insertion points
of these operators suggests that it is possible to violate winding
number conservation modifying the background charge.

\end{abstract}

\newpage

\tableofcontents

\newpage

\section{Introduction}

Several aspects of spacetime physics in string theory have been made
accessible due to 
the development of world-sheet methods. In particular, the significant
progress of 
algebraic techniques in rational conformal field theory (RCFT) has
been crucial to
improve our knowledge on string compactification and string
phenomenology. 
Nowadays the possibility of extending the systematic understanding
gained in RCFT to 
non-RCFT in order to describe non-compact backgrounds is under active
investigation 
(see \cite{schomerus} for a complete and comprehensive review).

In this article we consider the 
$SL(2, \mathbb{R})$ WZNW model, a non-RCFT describing string
propagation on three 
dimensional anti de Sitter spacetime ($AdS_3$). 
Beyond the interesting formal aspects involved in the study of  non-RCFT, this
model allows to address important physical issues such as 
curved backgrounds in string theory
 and the analysis of a genuine stringy regime of the AdS/CFT correspondence.

This theory has been
thoroughly studied in the 90's and its current status was
established in \cite{mo1}$ -$\cite {mo3} where several longstanding problems were
puzzled out.
The Hilbert space of physical states
contains all the Hermitian unitary representations of the universal
cover of 
$\widehat{SL}(2,\mathbb{R})$. These representations were found to be the
conventional unitary representations of the $SL(2,\mathbb R)$ current
algebra, namely, the principal continuous series 
$\widehat{\cal  C}_j^\alpha$ 
with $j=-\frac 12+i{\mathbb R}$ and $0\le\alpha<1$, and
the lowest- and highest-weight principal discrete series
$\hat{\cal D}_j^{\pm}$,
with $j\in {\mathbb R}$ and $-\frac 12<j<\frac{k-3}2$, and their spectral
flow images $\hat{\cal  C}_j^{\alpha, w}$ and $\hat{\cal D}_j^{\pm,
  w}$, respectively \cite{mo1}. Here $k$ is the level of the current
algebra and the spectral flow parameter $w$ is an
integer number that can
be identified with the winding number of  long strings stretched
close to the boundary of AdS$_3$.
A proof of the no-ghost theorem for this spectrum was given in
\cite{mo1} and verified by the exact calculation of the one loop
partition function of string theory on $AdS_3\times {\cal M}$  in \cite{mo2}, ${\cal
  M}$ being a compact space represented by a unitary CFT on the
worldsheet. 
To establish the
consistency of the full theory, interactions have to be considered in order 
to verify the closure  of the operator product expansion. Important progress  has been
achieved in the resolution of this problem in \cite{mo3} where some
correlation functions were computed and the structure of the
factorization of a 
four-point function was shown to agree with the physical spectrum of the
theory. However a complete proof of unitarity, involving arbitrary
spectral flow sectors, is still lacking. 

 Various correlators are known in this theory and its
euclidean version, the $H^+_3$ or $SL(2,\mathbb{C})/SU(2)$ 
coset model.
The structure constants
of the $H_3^+$ model were obtained in \cite{tesch1} using a
generalization of the bootstrap approach.
The analytic
continuation of these results to the $SL(2,\mathbb{R})$ WZNW model was discussed in
\cite{mo3} 
where two- and three-point functions involving spectral flowed states
were also computed. 
The methods followed in these works rely on the
invariance of correlation functions under the Lie and conformal
symmetries, which leads to the Knizhnik-Zamolodchikov (KZ) equations,
and they  also
 take advantage of the properties of the so-called spectral flow
operator, an auxiliary degenerate operator interpolating between
different $w$ sectors, which gives additional differential equations
to be satisfied by the correlators. Actually, a suitable number of these spectral flow
operators has to be inserted in the correlation functions in order to
transfer winding to the physical states. For instance 
the four-point function of three generic states and one
spectral flow operator was computed in \cite{mo3} in order to obtain
the one unit spectral flow three-point function \footnote{A well
  known feature of 
 string interactions is that 
 winding number conservation may be violated according to a precise
 pattern determined by 
the properties of $SL(2,\mathbb{R})$ representations \cite{mo3, zamo}.}.
The same method was used in
\cite{zamo, hmn} to compute the five-point function involving three generic states
and two spectral flow operators, leading to the spectral flow
conserving
three-point function of two $w=1$ and one $w=0$ states
\footnote{Recall that $w$ takes positive integer values in the
  $x-$basis  
\cite{mo3}.}.

The problem of constructing a complete set of solutions to the
KZ equations for generic four-point functions 
has not been solved yet. These equations 
cannot be reduced 
to a system of ordinary differential equations \cite{tesch2} as in
the case of the compact $SU(2)$ WZNW model \cite{zf} and thus the bootstrap
approach  does not lie on a firm rigorous
mathematical basis up to now.
Solutions have been found only as  formal power
series in  the cross-ratio on the worldsheet \cite{mo3, tesch2,
  mn}. 
Aiming to
explore an alternative method to compute four-point functions, 
in this article we consider the Coulomb gas formalism \footnote{
Another method, which exploits the relation
  between the
  $H_3^+$ and Liouville theories, was presented in \cite{ribault, hoso}.}. This approach
appears to be ad-hoc and
not very well grounded,  since there are no
singular
vectors in the unitary spectrum of $\widehat{SL}(2,\mathbb{R})$ while
it is well known that 
degenerate fields provide the  formal mathematical foundation
for the background charge method \cite{felder}. Some skepticism on
this procedure was raised in \cite{hhs},
where the Hilbert space of physical states of string theory on $AdS_3$ was
constructed as the 
BRST cohomology on the Fock spaces of free fields and
it was shown to differ from the
 spectrum determined in
\cite{mo1}. From a computational perspective,
one disadvantage of this method
is that it is implemented in the necessarily more
cumbersome $m-$basis. However, while only one unit spectral flow operators have
been constructed in the $x-$basis \cite{mo3}, it is only in the
$m-$basis 
that the spectral flow 
operation can be performed in arbitrary winding sectors so far.

The Coulomb gas  method was  used in reference \cite{becker} to compute
 two- and three-point functions \footnote{See \cite{otros} for Coulomb gas computations
 of SL(2) correlators involving states in admissible representations
 and \cite{fl} for more recent work on the related Liouville theory.}. It was later
 extended  to include winding and to obtain spectral flow
 non-conserving three-point
 functions 
in \cite{gn1,gn2}. 
The starting point of the
 procedure 
implemented in \cite{becker, gn2} involved three-point functions
 preserving winding 
number conservation with at least one highest-weight state, whereas
 the one unit 
spectral flow amplitudes included at least two highest-weight
 states. The highest-weight condition  was necessary in order to
 simplify the 
computation and manage to solve it explicitly. 
One of the
 difficulties
comes from the $\beta - \gamma$ ghost fields
 required by the 
Wakimoto realization and the fact that the underlying representations
 of 
$SL(2,\mathbb{R})$ are infinite dimensional. This leads to consider
 arbitrary numbers
 of screening currents and intricate ghost correlators. 
Another trouble
 of this approach 
arises in the values required by unitarity for the spin $j$ and the
 necessity to 
analytically continue to non-integer numbers of screening
 operators. The continuation 
performed in \cite{becker} leads to the exact results when the highest-weight condition 
is relaxed to include an arbitrary global descendant in the
 correlator \cite{satoh}, but so far it 
has not been shown that the most general case can be obtained in this
 way. Similarly, the spectral flow non-conserving three-point function
 computed in \cite{gn2} reproduces the exact result when it
 involves at least two  highest-weight states.

In this article,
we are 
able to compute the required expectation values of the $\beta -
\gamma$ system for generic three-point functions and for a particular
four-point function, using 
standard bosonization, the definition of the Vandermonde determinant
and  Schur
polynomials. This leads to multiple integral expressions for unrestricted
three-point 
functions which we manage to solve working out  Aomoto integral
\cite{aomoto} in
the complex plane, 
a new integral formula of  Dotsenko-Fateev type \cite{df1, df2}. We then make use of
monodromy 
invariance to analytically continue to non-integer numbers of
screening operators and 
show that the results obtained in this way are in full agreement with
those of 
\cite{mo3, tesch1, hmn} for generic spectral flow conserving three-point functions.
We also reproduce the computation of the one unit spectral flow 
three-point function performed in \cite{zamo, mo3} using free fields, and
again we show full agreement with the exact results. 
This problem involves the evaluation of a four-point function including one
spectral flow operator, which we manage to do extending the techniques
discussed above for the winding number conserving case. Thus we
 demonstrate the  complete coincidence between  correlation
functions computed in the
free field formalism and all the known  exact results. 
Additionally, we present an alternative method to
obtain winding number non-preserving $n-$point functions  
in the Coulomb gas approach  through well defined operators.
The independence of the result on the insertion points of these
operators suggests that it is possible to violate winding number
conservation modifying the background charge.
We prove that this is in fact the case for $n=3$.

This work is organized as follows. In order to set up our notation,
we include a brief 
review of the free field realization of $SL(2,\mathbb{R})$ in Section
2. The computation 
of  winding conserving three-point functions is given in Section
3. This section 
contains three parts: the resolution of the $\beta - \gamma$
correlator, the explicit 
evaluation of Aomoto integrals in the complex plane and finally, the
analytic 
continuation to non-integer numbers of screening operators.
Comparison with previous 
results obtained through the bootstrap approach
 is presented as well. In Section 4 we compute the
spectral flow non-conserving three-point function using the Coulomb gas
method following the
prescription developed in \cite{zamo, mo3}. We also present
a novel method to compute the winding number non-conserving $n-$point
functions and we  verify that it
reproduces the exact results for $n=3$.
Section 5 offers conclusions and a
discussion about 
 perspectives for the extension of the techniques used throughout the text
to generic
four-point 
functions.

\section{Free field realization of the $SL(2,\mathbb{R})$ WZNW model}

Mainly in order to set up our notation and conventions we will briefly review in this section the free field realization of the $SL(2,\mathbb{R})$ WZNW model and the computation of correlation functions in this approach.

\subsection{Wakimoto representation}

Bosonic string propagation in an $AdS_3$ background is described by a non-linear sigma model equivalent to a WZNW model on $SL(2,\mathbb{R})$ (actually, its Euclidean version on $SL(2,\mathbb{C})/SU(2)$). The action of the theory is given by
\begin{equation}
\label{sigmod}
S = \frac k{8\pi} \int d^2z ~ (\partial\phi \bar\partial\phi + e^{2\phi} \bar\partial\gamma  \partial\bar\gamma),
\end{equation}
where $\{\phi, \gamma , \bar\gamma \}$ are the Poincar\'e coordinates
of the (euclidean) 
$AdS_3$ spacetime and $k=l^2/l^2_s$, $l$ being related to the scalar
curvature of $AdS_3$
 as ${\cal R}=-2/l^2$ and $l_s$ being the fundamental string
 length. While $\phi$ is a 
real field, $\{\gamma , \bar\gamma \}$ are complex coordinates
parameterizing the 
boundary of $AdS_3$, which is located at $\phi\rightarrow\infty$.

Eq.~(\ref{sigmod}) can be obtained by integrating out the one-form
auxiliary fields 
$\beta$ and $\bar\beta$ in the following action:
\begin{equation}
\label{lag}
{\cal S} = \frac{1}{4\pi} \int d^2z \left[
  \partial\phi\bar\partial\phi - \frac{2}
{\alpha_+} \, {\cal R}^{(2)}\phi + \beta \bar\partial\gamma +
  \bar\beta\partial
\bar\gamma - \beta\bar\beta e^{- \frac{2}{\alpha_+} \, \phi}\right],
\end{equation}
where $\alpha_+=\sqrt{2(k-2)}$, ${\cal R}^{(2)}$ is the scalar
curvature of the 
worldsheet and $k$-dependent renormalization factors have been
included after 
quantizing \cite{david, distler}.

The linear dilaton term in (\ref{lag}) can be interpreted as the
effect of a 
background charge at infinity. In fact, the large $\phi$ region can be
explored 
treating the interaction, perturbatively, as a screening charge. In
this limit, 
the theory reduces to a free linear dilaton field $\phi$ and a free
$\beta - \gamma$ 
system with propagators given by the following OPE:
\begin{equation}
\label{propag1}
\phi(z) \phi(0) \sim - \ln z,
\end{equation}
\begin{equation}
\label{propag2}
\beta(z) \gamma (0) \sim \frac 1z .
\end{equation}
Similar expressions hold for the antiholomorphic content. In what
follows we will focus 
our discussion on the left moving component of the theory and we will
assume that all 
the steps go through to the right moving part as well, indicating the
left-right 
matching condition only if necessary.
 
Generic correlation functions would not be expected to be reliable in
 this approximation. Notice that the original action (\ref{sigmod})
 has a singularity in this limit. However we will demonstrate
 below that the free field computation of two- and three-point
 functions gives results in 
full agreement with the exact calculations. We will show that this is
 the case even when 
winding number conservation is violated and the computation of a 
four-point function
is necessarily involved.

The theory is invariant under the action of two copies of the
$SL(2,\mathbb{R})$ current 
algebra at level $k$. We will denote the currents generating this
algebra by $J^a$ with 
$a=\pm, 3$. These currents are given in the Wakimoto representation by
\begin{eqnarray}
\label{waki1}
J^+ = \beta, 
\end{eqnarray}
\begin{eqnarray}
\label{waki2}
J^3 = - \beta\gamma - \frac{\alpha_+}{2} \, \partial\phi, 
\end{eqnarray}
\begin{eqnarray}
\label{waki3}
J^-=\beta\gamma^2+\alpha_+\gamma\partial\phi + k \partial\gamma, 
\end{eqnarray}
and they verify the following OPE:
\begin{eqnarray}
\label{opeope1}
J^+(z)J^-(0) \sim \frac{k}{z^2} - \frac{2 J^3(0)}{z}, 
\end{eqnarray}
\begin{eqnarray}
\label{opeope2}
J^3(z) J^{\pm}(0) \sim \pm \frac{J^{\pm}(0)}{z}, 
\end{eqnarray}
\begin{eqnarray}
\label{opeope3}
J^3(z)J^3(0) \sim -\frac{k/2}{z^2}, 
\end{eqnarray}
in agreement with the $SL(2,\mathbb{R})$ current algebra at level $k$ 
commutation relations.

The Sugawara construction gives rise to the following energy-momentum tensor:
\begin{equation}
\label{sugawara}
T_{SL(2,\mathbb{R})} = -\frac 12 \, \partial\phi\partial\phi -
\frac{1}{\alpha_+} \, 
\partial ^2\phi + \beta\partial\gamma,
\end{equation}
which leads to a Virasoro algebra with central charge $c=3k/(k-2)$.

In the next section we will find that it is convenient to bosonize the 
$\beta - \gamma$ system. We do it in the standard way, i.e., 
after introducing free bosonic fields $u$ and $v$ with OPE
\begin{equation}
\label{bosonization}
u(z)u(0) = v(z)v(0) \sim - \ln z,
\end{equation}
we parametrize the $\beta - \gamma$ system as follows:
\begin{equation}
\label{bg1}
\beta=-i\partial v e^{-u+iv},
\end{equation}
\begin{equation}
\label{bg2}
\gamma=e^{u-iv} .
\end{equation}

The currents take now the following form:
\begin{eqnarray}
\label{bos1}
J^+ = - i\partial v e^{-u+iv},
\end{eqnarray}
\begin{eqnarray}
\label{bos2}
J^3 = \partial u - \frac{\alpha_+}{2} \, \partial \phi,
\end{eqnarray}
\begin{eqnarray}
\label{bos3}
J^- = e^{u-iv} [ (k-2) \partial u -i (k-1)\partial v + \alpha_+ \partial \phi ], 
\end{eqnarray}
while the energy-momentum tensor is given by
\begin{eqnarray}
\label{bos4}
T_{SL(2,\mathbb{R})}=-\frac 12 \,\partial u  \partial u -\frac 12 \,
\partial v 
\partial v -\frac 12 \, \partial \phi \partial\phi -\frac 12 \,
\partial ^2 u + 
\frac i2 \, \partial ^2 v - \frac{1}{\alpha_+} \, \partial^2\phi.
\end{eqnarray}

\subsection{Vertex operators and correlation functions}

The Hilbert space of physical states for this model was established in
\cite{mo1}. It contains all the Hermitian unitary representations of
the universal cover of $\widehat{SL}(2,\mathbb{R})$. The conventional unitary
representations of the current algebra, namely, the principal
continuous series $\hat{\cal C}_j^{\alpha,0}$ with $j=-1/2 +i
\lambda$, $\lambda \in \mathbb{R}$ and $0\le \alpha < 1$, and the
lowest- and highest-weight principal discrete series $\hat{\cal
  D}_j^{\pm,0}$, with $j\in \mathbb{R}$ are, of course, among those
included in the spectrum. In addition, their spectral flow images
$\hat{\cal C}_{j}^{\alpha, w}$ and $\hat{\cal D}_j^{\pm, w}$ must be
taken into account. These series are related to those 
with $w=0$ by the spectral flow automorphism of the current algebra,
i.e., the symmetry 
under which the modes of the currents transform as
\begin{equation}
\label{spfc}
J^{\pm}_n \rightarrow \tilde{J}^{\pm}_n = J^{\pm}_{n \pm w},
\end{equation}
\begin{equation}
\label{spfc2}
J^{3}_n \rightarrow \tilde{J}^{3}_n = J^{3}_{n} - \frac k2 \, w
\delta_{n,0} .
\end{equation}
As it was pointed out in \cite{mo1}, in standard models based on
compact Lie groups the
spectral flow does not generate new types of representations, but for
the 
$SL(2,\mathbb{R})$ WZNW model representations with different amounts
of $w$ are, in 
general, inequivalent. A relevant exception is the case of the series 
$\hat{\cal D}_j^{\pm, \mp 1}$ and $\hat{\cal D}_{\frac k2-2-j}^{\mp, 0}$.
The
equivalence of these representations 
 has an important consequence on the allowed values of
$j$, namely, 
for  discrete representations one has
\begin{eqnarray}
\label{allowed}
-\frac 12 < j < \frac{k-3}2 .
\end{eqnarray}

The full Hilbert space of the theory is, then,
\begin{eqnarray}
\label{hilbspace}
{\cal H} = \bigoplus_{w=-\infty}^{+\infty} \left\{
\left[\int_{-\frac{1}{2}}^{\frac{k-3}{2}} ~dj ~ \hat{\cal D}_j^{+,w} \otimes
  \hat{\cal D}_j^{+,w} \right] \oplus
\left[\int_{-\frac{1}{2}+i\mathbb{R}} dj \int_0^1 d\alpha \, \hat{\cal
    C}_j^{\alpha,w} \otimes \hat{\cal C}_j^{\alpha,w} \right] \right\} ,
\end{eqnarray}
where $\hat{\cal D}_j^{+,w}$ is the irreducible representation of the
$SL(2,\mathbb{R})$  current algebra generated from the highest-weight
state  $|j; w\rangle$ defined by
\begin{eqnarray}
\label{mal23}
J_{n+w}^+|j; w\rangle =J_{n-w-1}^-|j; w\rangle = J_n^3 |j; w\rangle =
0 ,\qquad (n=1, 2, \cdots )
\end{eqnarray}
\begin{eqnarray}
J_0^3|j; w\rangle = \left(j+\frac{k}{2} \,w\right) |j;w\rangle,
\end{eqnarray}
\begin{eqnarray}
\left[- (J_0^3 - \frac{k}{2} \, w)^2  + \frac{1}{2} \,\left(J_w^+
  J_{-w}^- + J_{-w}^- 
J_w^+ \right)\right] |j; w\rangle = -j(j+1) |j; w\rangle,
\end{eqnarray}
and $\hat{\cal C}_j^{\alpha,w}$ is generated from the state $|j,
\alpha; w\rangle$ 
obeying
\begin{eqnarray}
J_{n\pm w}^\pm |j, \alpha ; w\rangle = J_n^3 |j, \alpha; w\rangle = 0,
\end{eqnarray}
\begin{eqnarray}
J_0^3 |j, \alpha;w\rangle = \left(\alpha + \frac{k}{2}\,w\right)|j, \alpha; w \rangle,
\end{eqnarray}
\begin{eqnarray}
\label{mal29}
\left[- (J_0^3 - \frac{k}{2} \, w)^2  + \frac{1}{2} \,\left(J_w^+
  J_{-w}^- + J_{-w}^- 
J_w^+ \right)\right] |j, \alpha; w\rangle = -j(j+1) |j, \alpha; w\rangle.
\end{eqnarray}

A suitable representation of the vertex operators creating these
 states was introduced in the discrete light cone approach in
 \cite{hhs}. In terms of the fields $u$ and $v$,
these vertex operators can be written as
\begin{eqnarray}
\label{vcw}
V_{j,m,\bar m}^w= e^{(j-m-w)u(z)}e^{(j-\overline m-w)u(\bar z)}
e^{-i(j-m)v(z)} e^{-i(j-\overline m )v(\bar z)} e^{\left(\frac
  2{\alpha_+}j+ \frac {\alpha_+}2 w\right) \phi(z,\bar z)},
\end{eqnarray}
where we are assuming, as usual, that $(m -\bar m)$ is an integer number.

It is straightforward to check the following properties:
\begin{eqnarray}
\tilde{J}^\pm(z) V^w_{j,m,\bar m}(0) \sim \frac{\pm j - m}{z} V^w_{j,m\pm 1,\bar m}(0),
\end{eqnarray}
\begin{eqnarray}
\tilde{J}^3(z) V^w_{j,m,\bar m}(0) \sim \frac {m}z \, V^w_{j,m,\bar m}(0),
\end{eqnarray}
in agreement with (\ref{mal23})-(\ref{mal29}). Notice that 
the zero modes of the currents 
have been shifted according to the spectral
flow sector of the  vertex operators, i.e., the shift
(\ref{spfc})$-$(\ref{spfc2})  cancels the factors $z^{\pm w}$ and $kw/2z$ in the OPE.

The operators (\ref{vcw}) reduce to the well known vertices \cite{morozov}
\begin{eqnarray}
\label{vsw}
&& V_{j,m,\bar m}=\gamma^{j-m}\bar \gamma^{j-\bar m} e^{2j\phi/\alpha_+}
\end{eqnarray}
in the case $w=0$. For future reference, it is convenient to recall
here that these  operators appear when looking for the asymptotic
expressions of the  following normalizable operators in the
$SL(2,\mathbb{C})/SU(2)$  model
\begin{equation}
\label{xbasis}
\Phi_j(x,\bar x; z,\bar z) = \frac{1+2j}{\pi}\,\left (|\gamma - x|^2
e^{\phi/\alpha_+} +e^{-\phi/\alpha_+} \right )
^{2j},
\end{equation}
where $x, \bar x$ keep track of the $SL(2,\mathbb{C})$ quantum
numbers.  These operators, when transformed to the $m$-basis through
the following  Fourier integral:
\begin{equation}
\label{x-m}
\Phi_{j,m,\bar m}(z,\bar z) = \int d^2 x \, x^{j-m}\bar x^{j-\bar m}  
\Phi_{-1-j}(x,\bar x, z, \bar z) \quad ,
\end{equation}
give, in the limit $\phi\rightarrow \infty$,
\begin{equation}
\label{asymp}
\Phi_{j,m,\bar m}(z,\bar z) \sim V_{j,m,\bar m} + \frac{1+2j}{\pi} \,  
c_{m,\bar m}^{-1-j}V_{-1-j,m,\bar m},
\end{equation}
where 
\begin{eqnarray}
\label{cjmm}
c^j_{m,\bar m} = \pi \gamma(1+2j)\frac{\Gamma(-j+m)\Gamma(-j-\bar m)}
{\Gamma(1+j+m)\Gamma(1+j-\bar m)}
\end{eqnarray}
and
\begin{eqnarray}
\label{gammachica}
\gamma(x)=\frac{\Gamma(x)}{\Gamma(1-x)}.
\end{eqnarray}
While both terms in (\ref{asymp}) contribute if the state is in a
principal continuous series, Eq.~(\ref{asymp}) reduces to the vertex
(\ref{vsw}) when the state belongs to a highest or lowest-weight
representation. 

We can introduce  winding number in  (\ref{asymp})  and define
asymptotically
\begin{equation}
\label{asympw}
\Phi^w_{j,m,\bar m}(z,\bar z) \sim V^w_{j,m,\bar m} + \frac{1+2j}{\pi}
\,  
c_{m,\bar m}^{-1-j}V^w_{-1-j,m,\bar m}.
\end{equation}
Recall that (\ref{asymp}) is a classical
expression. The second term requires an additional factor to account
for quantum corrections \cite{tesch1} which is also necessary in (\ref{asympw}).

All relevant correlators will involve vertex operators like
(\ref{vcw}). 
In the Coulomb gas formalism it is mandatory to insert, in addition,
some operators in order to screen the charges of these vertices and
the background charge. 
The so-called screening operators in the $SL(2,\mathbb{R})$ WZNW model
are the 
following \cite{morozov}:
\begin{eqnarray}
{\cal S}_+ =  \int d^2 y \, \beta(y)\bar\beta(\bar y) e^{-2 \phi(y,\bar y)/\alpha_+},
\label{scree1}
\end{eqnarray}
\begin{eqnarray}
{\cal S}_- =  \int d^2 y \, \left [\beta(y)\bar\beta(\bar y)\right
]^{k-2} 
e^{-\alpha_+\phi(y,\bar y)},
\label{scree2}
\end{eqnarray}
where the screening currents are defined so that they have trivial OPE
  \footnote{Recall that there is a total derivative in the  OPE with
  $J^-$ which 
requires a careful treatment of contact terms \cite{becker}.}$^,$ 
\footnote{The relation between ${\cal S}_+$
  and the zero momentum mode of the dilaton was established in \cite{gk}.}
with $J^{3,\pm}$.

Summarizing, in the context of the Coulomb gas formalism one has to
compute
expectation values of the form 
\begin{equation}
\label{function22}
{\cal A}_n = \langle V_{j_1,m_1,\bar m_1}^{w_1}(z_1,\bar z_1)\cdots
V_{j_n,m_n,\bar m_n}^{w_n} (z_n,\bar z_n) \, {\cal S}_+^{s_+} \, {\cal
  S}_-^{s_-} \rangle ,
\end{equation}
where the number of screening operators is determined from the
following conservation laws \cite{becker}:
\begin{equation}
\sum_i\alpha_i^{\phi} = -\frac 2{\alpha_+}, \qquad
\qquad \sum_i\alpha_i^u=-1,\qquad\qquad \sum_i\alpha^v_i=i,
\end{equation}
 $\alpha^{\phi, u, v}$ being the charges of the fields $\phi, u, v$, respectively.
\footnote{
Spectral
  flow operators must be also considered if winding number
  conservation is not preserved. We will discuss this subject
  in section 4.}

Notice that ${\cal S}_+$ is the interaction term in (\ref{lag}), and
therefore computing amplitudes with $s_-=0$ is completely equivalent
to a perturbative expansion of order $s_+$ in the path integral
formalism. Moreover, as we have already seen, unitarity requires
$-\frac 12 < j < \frac {k-3}2$  for the principal discrete series or
$j=-\frac 12 + i\lambda$ with $\lambda \in \mathbb{R}$ for the
continuous ones. It is then necessary to consider  $s_+, s_- \notin
\mathbb{Z}^+$, i.e., after computing (\ref{function22}) for an integer
number of screening insertions it is necessary to continue this
function to non-integer values of $s_+$ and $s_-$. Actually, once this
generalization is allowed any correlator can be computed using only
one kind of screening operators. In what follows we will consider only
screening operators of type ${\cal S}_+$ and we will write ${\cal S}$ instead of ${\cal S}_+$, and $s$ instead of $s_+$.

\subsection{Two- and three-point amplitudes}

The two- and three-point functions in the $SL(2,\mathbb C)/SU(2)$
WZNW model 
were computed exactly in \cite{tesch1} and winding number was included in \cite{mo3}.
For future reference we quote here the results. 
\begin{eqnarray}
\label{twopoint}
&& \langle \Phi^{w}_{j,m,\bar m}(z)\Phi^{w'}_{j',m',\bar
  m'}(z')\rangle =  \delta^2(m+m') \, \delta_{w+w'} \,
  (z-z')^{-2\hat{\Delta}}
  (\overline{z}-\overline{z}')^{-2\overline{\hat{\Delta}}}  \nonumber \\
&& ~~~~~~~~~~~~~~~~~~~~~~ \qquad\qquad\qquad
\times ~ \left[ \delta(j+j'+1) +  B(1+j) c^{-1-j}_{m,\overline{m}} \,
  \delta(j-j') 
\right],
\end{eqnarray}
where
\begin{eqnarray}
\hat{\Delta} = \Delta(j)-wm-\frac{k}{4} w^2 =  -\frac{j(j+1)}{k-2}-wm-\frac{k}{4} w^2,
\label{Delta}
\end{eqnarray}
\begin{eqnarray}
\delta^2(m)\equiv \int d^2x \, x^{m-1} \overline{x}^{\overline{m}-1}=
4\pi^2  \delta(m+\overline{m}) \delta_{m-\overline{m}},
\label{deltam}
\end{eqnarray}
\begin{eqnarray}
B(j) =  - \frac{1}{\pi \rho} \, \frac{\nu^{1-2j}}{\gamma(\rho(1-2j))},
\label{2pf}
\end{eqnarray}
with $\nu$  given by
\begin{eqnarray}
\nu = -\frac{\pi}{\rho} \, \frac{1}{\gamma(-\rho)},
\label{nu}
\end{eqnarray}
and we have defined $\rho=-1/(k-2)$.
This value of $\nu$ was set in \cite{tesch1} by requiring consistency
between the two-  and three-point functions but its choice will not
affect the discussion  in the rest of the paper.

The Coulomb gas computation of the two-point function was performed in
reference \cite{becker} for  $m_i=\bar m_i$ and 
winding number was included in reference \cite{gn2} where the
expression (\ref{twopoint}) was obtained. 

It is well known that winding number conservation may be violated
in three and higher-point functions \cite{zamo, mo3}.
If we assume that winding number is conserved, i.e.,
\begin{eqnarray}
\label{conserv}
w_1+w_2+w_3=0 ,
\end{eqnarray}
 a generic three-point function has the following expression:
\begin{eqnarray}
\label{threepoint}
&& \left\langle \prod_{i=1}^3 \Phi^{w_i}_{j_i,m_i,\bar m_i}(z_i)
\right\rangle = \delta^2({\mbox{$\sum m_i$}}) \, C(1+j_i)
W(j_i,m_i,\overline{m}_i) \prod_{i<j}  |z_{ij}|^{-2\Delta_{ij}} ,
\end{eqnarray}
where $z_{ij}=z_i-z_j$, $\Delta_{12}=\Delta_1+\Delta_2-\Delta_3$, etc.

The function $C(j_i)$ is the three-point function of the primaries in
the $x$-basis  computed in \cite{tesch1}. It reads
\begin{eqnarray}
C(j_i) = - \frac{G(1-\sum j_i)G(-j_{12})G(-j_{13})G(-j_{23})}{2 \pi^2
  \nu^{j_1+j_2+j_3-1} \gamma\left(\frac{k-1}{k-2}\right) G(-1)
  G(1-2j_1) G(1-2j_2)  G(1-2j_3)},
\end{eqnarray}
where
\begin{eqnarray}
G(j)=(k-2)^{\frac{j(k-1-j)}{2(k-2)}} \, \Gamma_2(-j |1,k-2 ) \, \Gamma_2(k-1-j |1,k-2 ),
\end{eqnarray}
$\Gamma_2(x|1,w)$ being the Barnes double Gamma function and
$j_{12}=j_1+j_2-j_3$, etc.

The function $W(j_i,m_i,\overline{m}_i)$ is given by the following integral:
\begin{eqnarray}
&& W(j_i,m_i,\overline{m}_i) = \int d^2x_1\, d^2x_2\, x_1^{j_1-m_1}  
\overline{x}_1^{j_1-\overline{m}_1} x_2^{j_2-m_2}
\overline{x}_2^{j_2-\overline{m}_2} 
 \nonumber \\
&& ~~~~~~~~~~~~~~~~~~~~ \qquad\qquad
\times ~ |1-x_1|^{-2j_{13}-2} |1-x_2|^{-2j_{23}-2} |x_1-x_2|^{-2j_{12}-2}, 
\end{eqnarray}
which was computed in \cite{fukuda}.

The one unit spectral flow
 three-point function was first evaluated in \cite{mo3}. It is given by
\begin{eqnarray}
\label{3pointwind}
&& \left\langle \prod_{i=1}^3 \Phi^{w_i}_{j_i,m_i,\overline{m}_i}(z_i)
\right\rangle =  \delta^2(\mbox{$\sum m_i$}+\frac k2) \,
\frac{\tilde{C}(1+j_i)
  \tilde{W}(j_i,m_i,\overline{m}_i)}{\gamma(j_1+j_2+j_3+3-\frac k2)}   
\, \prod_{i<j} z_{ij}^{-\hat\Delta_{ij}}
\overline{z}_{ij}^{-\overline{\hat\Delta}_{ij}} ,
\end{eqnarray}
where now \footnote{Recall that our
  convention for $m$, and consequently for $w$, differs by a sign from
that in \cite{mo3}.}
\begin{eqnarray}
w_1+w_2+w_3=1,
\end{eqnarray}
\begin{eqnarray}
\tilde{C}(j_i) \sim B(j_1) C(\frac k2-j_1,j_2,j_3),
\end{eqnarray}
\begin{eqnarray}
\tilde{W}(j_i,m_i,\overline{m}_i)=\frac{\Gamma(1+j_1+m_1)}{\Gamma(-j_1-\overline{m}_1)}
\frac{\Gamma(1+j_2+m_2)}{\Gamma(-j_2-\overline{m}_2)}
\frac{\Gamma(1+j_3+\overline{m}_3)} {\Gamma(-j_3-m_3)},
\end{eqnarray}
and $\hat{\Delta}_{ij}$ equals $\Delta_{ij}$ with $\Delta_1$ replaced
by  $\Delta_1+m_1-k/4$.

The computation of 
three-point
functions of $w=0$ states
was performed in the Coulomb gas approach for correlators with at
least one
insertion in 
the principal discrete series in \cite{becker}. States in other
winding sectors 
were considered in \cite{gn2}, where winding number conserving three-point
functions with at least one
highest-weight 
state 
as well as winding number non-conserving three-point
functions with at least two highest-weight states were computed. As we have already 
mentioned the expressions in
the Coulomb gas 
approach agree with the exact ones. However,
the highest-weight condition greatly simplifies the computations and a
complete proof of the agreement is not available yet. So in the forthcoming
sections we
will develop techniques which will allow us to obtain the full
expressions (\ref{threepoint}) and (\ref{3pointwind}) using free fields.

\section {Spectral flow conserving three-point functions}

In this section we obtain the spectral flow conserving three-point
functions 
for generic states in the Coulomb gas approach.
As we have already said, we need to compute the following expression:
\begin{eqnarray}
&&{\cal A}_3 = \langle V_{j_1, m_1, \bar m_1 }^{w_1}(z_1) V_{j_2, m_2, 
\bar m_2 }^{w_2}(z_2)V_{j_3, m_3, \bar m_3}^{w_3}(z_3)) {\cal S}^s \rangle,
\label{3pts}
\end{eqnarray}
where 
$s$ is determined from the 
 conservation laws, which in this case lead to
\begin{eqnarray}
&& \sum_{i=1}^3 j_i+1=s,
\label{conlaw1}
\end{eqnarray}
\begin{eqnarray}
&&\sum_{i=0}^3 m_i=\sum_{i=0}^3\bar m_i=0,
\label{conlaw2}
\end{eqnarray}
\begin{eqnarray}
&& \sum_{i=1}^3 w_i = 0.
\label{conlaw3}
\end{eqnarray}

The expectation value (\ref{3pts}) involves a $\phi$-dependent part
and a ghost correlator, i.e., a 
$u$-$v$-dependent part. The first one is trivial and gives  the
following 
Coulomb integrals:
\begin{equation}
\label{3fi}
\Gamma(-s)\int \prod_{i=1}^s d^2
y_i \prod_{k=1}^3|z_k-y_i|^{-4\rho j_k+2 w_k}
\prod_{i<j}^s |y_i-y_j|^{4\rho},
\end{equation}
where the factor $\Gamma(-s)$ is the contribution of the zero
modes. The ghost correlator does depend upon the $y_i$'s 
and it must be included into the integrand of (\ref{3fi}). It can be written as
\begin{eqnarray}
\langle \prod_{l=1}^3 e^{(j_l-m_l-w_l)u(z_l)}
\prod_{i=1}^s
  e^{-u(y_i)} \rangle 
 \langle \prod_{l=1}^3 e^{-i(j_l-m_l)v(z_l)}
\prod_{i=1}^s \partial_i e^{iv(y_i)} \rangle = 
 \prod_{i=1}^s \prod_{l=1}^3
  (z_l-y_i)^{-w_l} \mathcal{P}^{-1} \partial_1 \cdots 
\partial_s \mathcal{P},
\label{c22}
\end{eqnarray}
where
\begin{equation}
\mathcal{P}=\prod_{i=1}^s \prod_{l=1}^3(z_l-y_i)^{m_l-j_l}\prod_{i<j}(y_i-y_j).
\label{sergio322}
\end{equation}
Here we have only quoted the holomorphic component; the
antiholomorphic part has the same form, with the replacement
$m_l\rightarrow \overline{m}_l$.

Expressions (\ref{3fi})$-$(\ref{sergio322}) simplify when
$(z_1,z_2,z_3)=(0,1,+\infty)$. Indeed, in this case 
(\ref{3fi}) reads
\begin{equation}
\label{fi}
\Gamma(-s)\int \prod_{i=1}^s d^2
y_i |y_i|^{-4\rho j_1+2 w_1}|1-y_i|^{-4\rho j_2+2 w_2}
\prod_{i<j}^s |y_i-y_j|^{4\rho},
\end{equation}
and (\ref{sergio322}) is
\begin{equation}
\mathcal{P}=\prod_{i=1}^s y_i^{m_1-j_1}(1-y_i)^{m_2-j_2}\prod_{i<j}(y_i-y_j).
\label{sergio22}
\end{equation}

Notice that the dependence on the winding number cancels when
combining (\ref{fi}) and (\ref{c22}). This is an explicit verification
of the observation made in \cite{mo3, ribault} that the
coordinate independent part of the correlation
functions in the $m-$basis depend on the sum $\sum_iw_i$.
 
The ghost contribution to the three-point function was evaluated in
\cite{becker}  for the specific case in which all
$m_l=\overline{m}_l$,  $w_l=0$  and the state created by  
$V_{j_1, m_1, \bar m_1 }^{w_1}$ is a highest-weight state, namely, 
$m_1 = \bar m_1 = j_1$. We now compute it for three generic states. 

\subsection{Evaluation of the ghost correlator}

Equation (\ref{sergio22}) can be rewritten, up to an irrelevant sign, 
in terms of the Vandermonde determinant, i.e.,
\begin{equation}
\label{pe}
{\mathcal P}=\left[\prod_{i=1}^s y_i^{m_1-j_1}(1-y_i)^{m_2-j_2}\right]
\det 
\left(y_i^{j-1}\right),
\end{equation}
and then 
\begin{equation}
{\mathcal P}= \det\left[y_i^{j-j_1+m_1-1}(1-y_i)^{m_2-j_2}\right].
\end{equation}

Since each row in this determinant depends upon a single variable, the
multiple 
derivatives in equation (\ref{c22}) can be computed row by row with
only one derivation, 
namely,
\begin{eqnarray}
\partial_1 \cdots \partial_s \mathcal{P} = \det\left\{\partial_i \left 
[ y_i^{j-j_1+m_1-1} (1-y_i)^{m_2-j_2}\right ] \right \}.
\end{eqnarray}
Performing the derivatives in this last determinant we get 
\begin{eqnarray}
\label{det1}
\partial_1 \cdots \partial_s \mathcal{P} = \left[ \prod_{i=1}^s
  y_i^{m_1-j_1-1} 
(1-y_i)^{m_2-j_2-1} \right] \det\left(\ell^0_j y_i^{j-1} + \ell^1_j y_i^{j}\right),
\end{eqnarray}
with
\begin{eqnarray}
\ell^0_j= j-j_1+m_1-1,
\end{eqnarray}
\begin{eqnarray}
\ell^1_j= 1-j+j_1-m_1+j_2-m_2.
\end{eqnarray}
From here we obtain
\begin{eqnarray}
\mathcal{P}^{-1}\partial_1 \cdots \partial_s \mathcal{P} = 
\left[ \prod_{i=1}^s y_i^{-1} (1-y_i)^{-1} \right] \frac{\det\left(
  \ell^0_j y_i^{j-1} 
+ \ell^1_j y_i^{j}\right)}{\det(y_i^{j-1})}.
\end{eqnarray}

The determinant in the numerator of this equation  may be computed
performing the 
multiple distributions and noticing that the only non-vanishing
contributions come 
from those determinants in which the columns have all different powers. Therefore,
\begin{eqnarray}
\det\left( \ell^0_j y_i^{j-1} + \ell^1_j y_i^{j}\right) 
= \sum_{n=0}^s \left[ \prod_{k=1}^{s} \ell^{\lambda^n_{s+1-k}}_{k}
  \right] 
\det(y_i^{j-1+\lambda^n_{s+1-j}}),
\end{eqnarray}
where for every $n=1,2,\dots,s$ we have introduced the partition
\begin{eqnarray}
\lambda^n=(\underbrace{1,1,\dots,1}_{\mbox{\scriptsize $n$ entries}}),
\end{eqnarray}
and we have set $\lambda^{n}_k=0$ when $k>n$ and $\lambda^0=0$.

Using that
\begin{eqnarray}
\ell^0_1\cdots \ell^0_{s-n} 
= \frac{\Gamma(m_1-j_1+s-n)}{\Gamma(m_1-j_1)},\label{eles}
\end{eqnarray}
and
\begin{eqnarray}
\ell^1_{s-n+1} \cdots \ell^1_s 
= \frac{\Gamma(n+1-s+j_1-m_1+j_2-m_2)}{\Gamma(1-s+j_1-m_1+j_2-m_2)},
\end{eqnarray}
we finally obtain
\begin{eqnarray}
 \mathcal{P}^{-1} \partial_1 \cdots \partial_s \mathcal{P} &=& 
\left[ \prod_{i=1}^s y_i^{-1} (1-y_i)^{-1} \right] \sum_{n=0}^s 
\frac{\Gamma(m_1-j_1+s-n)}{\Gamma(m_1-j_1)}  \nonumber \\
&& ~~~~~ ~~~ \qquad\qquad
\times ~
\frac{\Gamma(n+1-s+j_1-m_1+j_2-m_2)}{\Gamma(1-s+j_1-m_1+j_2-m_2)} \, 
\frac{\det \left(y_i^{j-1+\lambda^n_{s+1-j}}  \right)}{\det(y_i^{j-1})}. \label{pes}
\end{eqnarray}

Notice that (\ref{eles}) vanishes when $m_1=j_1$, i.e., when 
$V^{w_1}_{j_1,m_1,\bar m_1}$ creates a highest-weight state, unless
$n=s$. 
In that case only one term of the above sum survives and the
computation reduces to 
the one performed in \cite{becker} with the following result:
\begin{eqnarray}
\mathcal{P}^{-1} \partial_1 \cdots \partial_s \mathcal{P} =
\frac{\Gamma(1-j_2-m_2)}
{\Gamma(-j_3+m_3)} \prod_{i=1}^s (1-y_i)^{-1},
\end{eqnarray}
where the conservation laws have been used
. 

The quotient of determinants in (\ref{pes}) is the Schur polynomial 
associated with the partition $\lambda^n$.
Actually it reduces to the elementary symmetric polynomial
\begin{eqnarray}
\label{elementary}
\alpha^s_n(y_1,\dots,y_s) = 
\frac 1{n!(s-n)!}\sum_{\sigma_s}\prod_{i=1}^n y_{\sigma_s(i)},
\end{eqnarray}
since $\lambda^n$ is the minimal partition of degree $n$. 
In the expression above the sum runs over all permutations of degree
$s$ and we have 
defined $\alpha_0^s=1$. We thus may write
\begin{eqnarray}
\mathcal{P}^{-1} \partial_1 \cdots \partial_s \mathcal{P} &=& \left[
    \prod_{i=1}^s 
y_i^{-1} (1-y_i)^{-1} \right]
    \frac{\Gamma(1+j_1-m_1)}{\Gamma(-j_3+m_3)}  
\nonumber \\
&& ~~~~~ ~~~~~ ~~~ \times ~\sum_{n=0}^s (-1)^{s+n}
    \frac{\Gamma(-j_3+m_3+n)}
{\Gamma(1-s+j_1-m_1+n)} \, \alpha^s_n(y_1,\dots,y_s),
\end{eqnarray}
where we have used the conservation laws and, repeatedly, the relation
\begin{eqnarray}
\Gamma(1+z-n) = (-1)^n\frac{\Gamma(1+z)\Gamma(-z)}{\Gamma(n-z)},
\end{eqnarray}
which holds for  $n \in \mathbb{N}$.

Therefore, the three-point function is given by
\begin{eqnarray}
&& {\cal A}_3 = \Gamma(-s) \,
  \frac{\Gamma(1+j_1-m_1)}{\Gamma(-j_3+m_3)} 
\frac{\Gamma(1+j_1-\overline{m}_1)}{\Gamma(-j_3+\overline{m}_3)} 
\sum_{n,\overline{n}=0}^s
  \frac{\Gamma(-j_3+m_3+n)}{\Gamma(1-s+j_1-m_1+n)} 
\nonumber \\
&& ~~~~~ ~~~~ \qquad\qquad \times ~\frac{\Gamma(-j_3+\overline{m}_3+\overline{n})}
{\Gamma(1-s+j_1-\overline{m}_1+\overline{n})} \, (-1)^{n+\overline{n}} 
\mathcal{A}_s^{n,\overline{n}}(-2j_1\rho,-2j_2\rho,\rho),
\label{a3}
\end{eqnarray}
where we have introduced the following integrals:
\begin{eqnarray}
 \mathcal{A}_s^{n,\overline{n}}(-2j_1\rho,-2j_2\rho,\rho) &=&
\int \prod_{i=1}^s d^2y_i \prod_{i=1}^s |y_i|^{-4j_1\rho-2}
|1-y_i|^{-4j_2\rho-2} \nonumber \\
&& ~~~~~ ~~~~~ ~~~~~ ~~~~~ ~~ \times ~
\prod_{i<j}^s |y_i-y_j|^{4\rho} \, \alpha^s_n(y_1,\dots,y_s) \, 
\alpha^s_{\overline{n}}(\overline{y}_1,\dots, \overline{y}_s).
\label{df5dfdf}
\end{eqnarray}

In the next subsection we will perform these multiple integrations.

\subsection{Aomoto integrals in the complex plane}

We need to calculate the following expression:
\begin{eqnarray}
&& \mathcal{A}_s^{n,\overline{n}} = \int d^2y \prod_{i=1}^s
  |y_i|^{2a-2} 
|1-y_i|^{2b-2} \prod_{i<j}^s |y_i-y_j|^{4\rho} \, \alpha^s_n(y) \, 
\alpha^s_{\overline{n}}(\overline{y}),
\label{df5}
\end{eqnarray}
where we are writing $\alpha^s_n(y)$ instead of
$\alpha^s_n(y_1,\dots,y_s)$ and 
$d^2y$ instead of $\prod_{l=1}^s d^2 y_i$. We will do it by using
the results found by Aomoto 
\cite{aomoto} and a contour manipulation similar to the one discussed
in \cite{df2}. 
We present the details here and include several useful formulas in the Appendix.

Let us start transforming these complex integrals into multiple
contour integrals 
following \cite{dotsenko}. It is convenient to introduce the set of
real variables
$y_l=u_l + i v_l$ for $l=1,2,\dots,s$, in terms of which
$\mathcal{A}_s^{n,\overline n}
(a,b,\rho)$ takes the form
\begin{eqnarray}
 \mathcal{A}_s^{n,\overline{n}} &=& \int du \, dv \, 
\prod_{l=1}^s (u_l^2 +  v_l^2)^{a-1} ((1-u_l)^2+v_l^2)^{b-1}  \nonumber \\ 
&& ~~~~\qquad\qquad
 \times ~\prod_{l<m}^s ((u_l-u_m)^2+(v_l-v_m)^2)^{2\rho} \alpha^s_n(u
 + i v) 
\alpha^s_{\overline{n}}(u - i v).
\end{eqnarray}
The integrations are now performed on the real axis.

Next, we analytically continue the contours of integration of the
$v$'s and we shift 
them close to the imaginary axis, i.e., we perform the change of
variables: 
$v_l\rightarrow -i{\rm exp}(-2i\epsilon)v_l$, where $\epsilon$ is a
vanishingly 
small positive number. Thus we may now rewrite the previous expression as
\begin{eqnarray}
 \mathcal{A}_s^{n,\overline{n}} &=& (-i)^s \int  du \, dv \, 
\prod_{l=1}^s (u_l^2 -  e^{-4 i \epsilon} v_l^2)^{a-1} 
((1-u_l)^2-e^{-4i\epsilon} v_l^2)^{b-1}  \nonumber \\
&& \qquad\qquad\times ~ \prod_{l<m}^s ((u_l-u_m)^2- e^{-4i\epsilon} 
(v_l-v_m)^2)^{2\rho} \alpha^s_n(u + e^{-2i \epsilon}
v)\alpha^s_{\overline{n}}
(u -e^{-2 i\epsilon} v).
\end{eqnarray}
An irrelevant phase factor of the form $e^{-2is\epsilon}$ was omitted
in the previous 
equation. 

The additional  change of integration variables $z_l=u_l+v_l$ and
$w_l=u_l-v_l$
gives the following form for the integral (\ref{df5}):
\begin{eqnarray}
\mathcal{A}_s^{n,\overline{n}} &=& \int dz \, dw \, 
\prod_{l=1}^s (z_l - i \epsilon (z_l-w_l))^{a-1} (w_l + i \epsilon
(z_l-w_l))^{a-1} (1-z_l+i\epsilon (z_l-w_l))^{b-1} 
\nonumber \\ 
&& \qquad\times ~ 
(1-w_l-i\epsilon 
(z_l-w_l))^{b-1} \prod_{l<m}^s (z_l-z_m- 
-i\epsilon(z_l-w_l+z_m-w_m))^{2\rho} \nonumber\\
&& \qquad\times ~ (w_l-w_m+i\epsilon(z_l-w_l+z_m-w_m))^{2\rho} 
\alpha^s_n(z- i \epsilon (z-w)) \alpha^s_{\overline{n}}(w + i\epsilon (z- w)),
\label{df2}
\end{eqnarray}
where we have written $dz \, dw$ in place of 
$\prod_{l=1}^s  \left (-idz_l \,dw_l/ 2\right )$.

After performing the limit $\epsilon\rightarrow 0^+$ this last double
integral 
factorizes as a product of two single contour integrals. The
$\epsilon$-dependent 
terms determine how the integration contours should be deformed in
order to avoid the 
singularities at $0$ and $1$ and to keep them away from each
other. The order in which 
the integrations in the $z$'s are to be made define the way in which
the integration 
contours corresponding to the $w$'s should be arranged: if $z_i < z_j$
then the contour 
of $w_i$ must lie below the one of $w_j$. Then, the limit $\epsilon
\rightarrow 0^+$ 
must be performed. See \cite{df2} for more details on this kind of
manipulation of 
complex contours.

The integral (\ref{df2}) can therefore be written as
\begin{eqnarray}
\mathcal{A}_s^{n,\overline{n}} = \left (-\frac i2 \right )^s
\sum_{\sigma} A^n_{\sigma}
(a, b, \rho) \times J^{\overline n}_{\sigma}(a, b, \rho),
\end{eqnarray}
where $\sigma$ runs over all orderings of the $z$'s. $A^n_{\sigma}(a,
b, \rho)$ denotes 
the integrals over the $z$'s ordered according to $\sigma$ and 
$J^{\overline n}_{\sigma}(a, b, \rho)$ denotes the contour integrals
of the $w$'s 
with the prescription on the contours that follows, as we have described, from $\sigma$.

If one of the $z$'s is not in the interval $(0,1)$, then at least one
of the contours of 
the $w$'s can be deformed to infinity and thus the integral
vanishes. On the other hand, 
since $\alpha_n^s(z_1,\cdots , z_n)$ is a symmetric polynomial, the
integration limits
 in $A_\sigma ^n(a,b,\rho)$ can be freely set to $0$ and $1$, showing
 that actually 
$A_\sigma ^n(a,b,\rho)$ does not depend on $\sigma$ but only on
 $s$. It is given by the 
following  integral:
\begin{eqnarray}
A_s ^n(a,b,\rho) = \int\limits_{0}^{1} dz_1 \cdots \int\limits_{0}^{1}
dz_s 
\prod_{i=1}^s z_i^{a-1} (1-z_i)^{b-1} \prod_{i<j}^s (z_i-z_j)^{2\rho}  \alpha^s_n(z).
\end{eqnarray}
This is Aomoto integral of order $n$ whose explicit expression we recall in the Appendix.

The integral $J_{\sigma}^{\overline{n}}(a,b,\rho)$ is given by
\begin{eqnarray}
J_{\sigma}^{\overline{n}}(a,b,\rho) = \int\limits_{\Lambda_1} dw_1
\cdots 
\int\limits_{\Lambda_s} dw_s \prod_{i=1}^s w_i^{a-1} (1-w_i)^{b-1}
\prod_{i<j} 
(w_i-w_j)^{2\rho} \alpha^s_{\overline{n}}(w),
\label{dff3}
\end{eqnarray}
where every contour $\Lambda_i$ comes from $-\infty$ in the lower half
complex plane 
and goes to $+\infty$ in the upper half complex plane crossing the
real line in $(0,1)$ 
with none of the contours intersecting another one. Notice that all
these contours 
can be deformed in such a way that each integration can be performed
from $+\infty$ 
in the lower half complex plane to $+\infty$ in the upper half complex
plane encircling 
the singularity at $1$ clockwise and the contours do not intersect with each other.

The factor $\prod_{i<j} (w_i-w_j)^{2\rho}$  is defined so that if all
the $w_i$ are placed on the real axis and decreasingly ordered, then
the phases of the multivalued products are all equal to zero. When the
variable $w_i$ is continued along its contour and is taken around some
other point, say $w_j$, in such a way that the contour of $w_i$ goes
above $w_j$, the product $\prod_{i<j} (w_i-w_j)^{2\rho}$ gets an
additional phase factor 
$\exp({-2\pi i \rho})$. 
It follows that the integral (\ref{dff3}) can be identified, up to
phase factors 
depending neither on $n$ nor on $\overline{n}$, with 
\begin{eqnarray}
\label{90}
J_{s}^{\overline{n}}(a,b,\rho) = \int\limits_{1}^{\infty} dw_1 \cdots 
\int\limits_{1}^{\infty} dw_s \prod_{i=1}^s w_i^{a-1} (1-w_i)^{b-1}
\prod_{i<j} 
|w_i-w_j|^{2\rho} \alpha^s_{\overline{n}}(w),
\end{eqnarray}
which is, again,  independent of $\sigma$.

Performing the change of variables $w_i \rightarrow y_i=1/w_i$ in
(\ref{90}) and using 
the inversion identity
\begin{eqnarray}
\alpha^s_{s-\overline{n}}(1/y_1,\dots,1/y_s) = \left[\prod_{i=1}^s
  y_i^{-1} \right]
\alpha^s_{\overline{n}} (y_1,\dots,y_s),
\end{eqnarray}
one gets that $J_{s}^{\overline{n}}(a,b,\rho)$ equals
$A_s^{s-\overline{n}}
(-a-b-2\rho(s-1),b,\rho)$. 

Summarizing, we have found that
\begin{eqnarray}
\mathcal{A}_s^{n,\overline{n}} = C A_s^n(a,b,\rho)
A_s^{s-\overline{n}}
(-a-b-2\rho(s-1),b,\rho), 
\end{eqnarray}
where the factor $C$  depends neither on $n$ nor on
$\overline{n}$. Actually, the independence of this factor on these
parameters allows us to read it from Dotsenko-Fateev integral 
(formula (B.9) of \cite{df2}) that corresponds to the case $n=\overline{n}=0$.

We finally get
\begin{eqnarray}
\label{correpta}
 \mathcal{A}_s^{n,\overline{n}} &=& \left( 
\begin{array}{c}
	s \\ n
\end{array}
\right) \left( 
\begin{array}{c}
	s \\ \overline{n}
\end{array}
\right) \frac{\Gamma(\alpha+s)\Gamma(2-\alpha-\beta-2s) }
{ \Gamma(1-\alpha-s)\Gamma(\alpha+\beta+2s-1) } \frac{\Gamma(\alpha+\beta+2s-n-1) 
\Gamma(1-\alpha-s+\overline{n})}{\Gamma(\alpha+s-n) 
\Gamma(2-\alpha-\beta-2s+\overline{n})} \times \mathcal{S}_s, \nonumber\\
\label{com}
\end{eqnarray}
where $\alpha=a/\rho$, $\beta=b/\rho$ and $\mathcal{S}_s$ is Dotsenko-Fateev integral.

Similarly as in the real case, all Aomoto integrals of definite order
in the complex 
plane can be placed together in a single expression (see the
Appendix). In fact, let us 
consider the integral
\begin{eqnarray}
\mathcal{A}_s(z,\overline{z}) = \int d^2y \prod_{i=1}^s |y_i|^{2a-2}
|1-y_i|^{2b-2} 
|z-y_i|^{2} \prod_{i<j}^s |y_i-y_j|^{4\rho}.
\label{df77}
\end{eqnarray}
This  is a polynomial in $z$ and $\overline{z}$ whose coefficients
precisely are, up to 
a phase, those integrals that we have evaluated. Replacing their
expressions into 
(\ref{df77}) we get
\begin{eqnarray}
\mathcal{A}_s(z,\overline{z}) & = & \frac{1}{4^s} \, \left| 
{\bar P}^{\alpha-1,\beta-1}_s(1-2z) \right|^2 \, \mathcal{S}_s,
\end{eqnarray}
where $\bar P^{\alpha, \beta}_s$ are the monic Jacobi polynomials
whose definition we  
recall in equation (\ref{ec3}) in the Appendix. Equivalently
we may write
\begin{eqnarray}
&& \mathcal{A}_s(z,\overline{z}) =
  \frac{\gamma(\alpha+s)}{\gamma(\alpha)} \,
  \frac{\gamma(\alpha+\beta+s-1)}{\gamma(\alpha+\beta+2s-1)} 
\left| {}_2F_1\left(-s,\alpha+\beta+s-1;\alpha;z\right) \right|^2 \,
  \mathcal{S}_s .
\label{jaco}
\end{eqnarray}

We may now go back to (\ref{a3}) and perform the sums. 

\subsection{Analytic continuation}

Let us recall that we have to evaluate the following sums:
\begin{eqnarray}
{\cal A}_3 &=& \Gamma(-s) \,
\frac{\Gamma(1+j_1-m_1)}{\Gamma(-j_3+m_3)} 
\frac{\Gamma(1+j_1-\overline{m}_1)}{\Gamma(-j_3+\overline{m}_3)}
\sum_{n,
\overline{n}=0}^s \frac{\Gamma(-j_3+m_3+n)}{\Gamma(1-s+j_1-m_1+n)}  \nonumber \\
&& ~~~~~ ~~~~ \qquad\qquad\times~
\frac{\Gamma(-j_3+\overline{m}_3+\overline{n})}
{\Gamma(1-s+j_1-\overline{m}_1+\overline{n})} \, (-1)^{n+\overline{n}} 
\mathcal{A}_s^{n,\overline{n}}(-2j_1\rho,-2j_2\rho,\rho),
\label{a3bis}
\end{eqnarray}
where $\mathcal{A}_s^{n,\overline{n}}(-2j_1\rho,-2j_2\rho,\rho)$ is
given in (\ref{com}) 
with the replacements $a=-2j_1\rho$ and $b=-2j_2\rho$.

First note that the combinatorial numbers in (\ref{com}) allow us to
freely extend the 
sum to $\infty$ and write the result in terms of a generalized
hypergeometric 
function. In fact, this leads to the result found in reference \cite{becker}, namely,
\begin{eqnarray}
{\cal A}^{j_1j_2j_3}_{m_1m_2m_3} = {\cal C} \overline{\cal C} {\cal I}(j_1,j_2,j_3,\rho),
\label{a3becker}
\end{eqnarray}
where
\begin{eqnarray}
{\cal C} = \frac{\Gamma(-2j_3)\Gamma(1+j_2+m_2)
    \Gamma(1+j_2+j_3+m_1)}
{\Gamma(-j_3-m_3) \Gamma(-j_1+m_1) \Gamma(1-m_1-j_3+j_2)} 
{_3 F_2}\left [ \left. {\begin{array}{c}
-j_3+m_3, -m_1-j_1, 1-m_1+j_1 \\ 
-m_1-j_2-j_3, 1-m_1+j_2-j_3
\end{array}} \right| 1 \right ].
\end{eqnarray}
$\overline{\cal C}$ is the same expression with the
replacement $m_i \rightarrow \overline m_i$ and 
${\cal I}(j_1,j_2,j_3,\rho) = \Gamma(-s){\cal
  S}_s(-2j_1\rho,-2j_2\rho,\rho)$.

It was pointed out by Y. Satoh \cite{satoh} that this result agrees,
up to a phase, with the integral transform to the $m-$basis of the
three-point function computed by J. Teschner \cite{tesch1} in the
$x-$basis, only when the amplitude involves at least one state from
the discrete representation. This seems to be a natural conclusion of
the procedure implemented in reference \cite{becker}, where the
starting point is a three-point function containing one highest-weight
state, and then acting with the lowering operator $J^-_0$ and using
the Baker-Campbell-Hausdorff formula, the dependence on $m_1=j_1$ is
changed in an integer amount. Indeed, the sums leading to
(\ref{a3becker}) in reference \cite{becker} sweep the highest-weight
representation. But we are considering three generic states here and
we arrive at the same result. The common assumption in both procedures
though is that the number of screening operators $s$ is an integer
number. The analytic continuation to non-integer values of $s$ was
performed in \cite{becker} for the particular case of on-shell
tachyons in the $SL(2,\mathbb{R})/U(1)$ coset model representing the
two dimensional black hole. We now use monodromy invariance to
analytically continue the dependence on $s$ and then show that this
leads to the complete exact result for generic three-point functions.

Let us start by noticing that equation (\ref{com}) may be recast as follows:
\begin{eqnarray}
\label{ser1}
\mathcal{A}_s^{n,\overline{n}}(-2j_1\rho,-2j_2\rho,\rho) =
\frac{(-1)^{n+\overline{n}}}{\pi\Gamma(0)}  \, \frac{\gamma(s-2j_1
  -2j_2 -1 )} {\gamma(2s-2j_1-2j_2-1)} 
\frac{\gamma(-2j_1+s)}{\gamma(-2j_1)} \,
\mathcal{S}_s(-2j_1\rho,-2j_2\rho,\rho) 
I(n,\overline{n}, s),
\end{eqnarray}
where 
\begin{eqnarray}
I(n,\overline n, s) =
\frac{\pi\gamma(-2j_1)}{\gamma(-s)\gamma(-j_{12})} 
\frac{\Gamma(n-s)\Gamma(-\bar n) \Gamma(1+2j_{3}-\bar n)
  \Gamma(-j_{23}+n)}
{\Gamma(1+s+\bar n)\Gamma(1+n)\Gamma(-2j_3+n) \Gamma(1+j_{23}-\bar n)}.
\end{eqnarray}
This combination of $\Gamma$-functions can be rewritten using  the
formula derived 
in reference \cite{mo3}, namely,
\begin{eqnarray}
\label{ser2}
&& I(n,\overline{n}, s) = \int d^2u u^{-s+n-1} \overline{u}^{-s+\overline{n}
-1} \left( |F(-s,s-2j_1 -2j_2 -1,-2j_1;u)|^2 \right. \nonumber \\
&& ~~~~~~~~~~~~~~ ~~~~ \qquad\qquad\qquad\qquad
+ \left. \lambda |u^{1+2j_1} F(s-2j_2 ,1-s+2j_1,2+2j_1;u)|^2\right),
\label{malda}
\end{eqnarray}
with
\begin{eqnarray}
\label{ser3}
\lambda= -\frac{\gamma(-2j_1)^2 \gamma(1-s+2j_1) \gamma( s-2j_2 )}
{(1+2j_1)^2 \gamma(-s) \gamma(s-2j_1 -2j_2 -1)}\quad .\nonumber
\end{eqnarray}
Here $F$ denotes the hypergeometric function ${}_2F_1$.

The factor $\gamma(-s)$ in the denominator of $\lambda$
diverges for integer 
$s$. Therefore the second term in the integral (\ref{ser2}) would not
contribute 
in this case. However, as discussed in \cite{mo3}, the sum of
hypergeometric functions 
in (\ref{ser2}) is the unique monodromy invariant combination. So we
claim that the full 
expression has to be used in order to properly analytically continue
$s$ to non-integer 
values.

The requirement of monodromy invariance could seem unnatural in
the context of  three-point functions since 
the simplest 
non-trivial case among multipoint correlators that involves invariance
under the action 
of the monodromy group is the four-point function. 
A way to understand this follows
from the fact 
that any three-point function can be obtained by acting on a
particular well defined 
four-point-like function with a suitable differential
operator. Indeed, from (\ref{a3}) 
and (\ref{jaco}) it follows that
\begin{eqnarray}
	\mathcal{A}_3 = \Gamma(-s) \mathcal{O} \,
	\overline{\mathcal{O}} 
\mathcal{A}_s(z,\overline{z})\large |_{z,\overline z =0},
\end{eqnarray}
where
\begin{eqnarray}
	\mathcal{O} =
	\frac{\Gamma(1+j_1-m_1)}{\Gamma(-j_3+m_3)\Gamma(-s)
	  \Gamma(1+s)} 
\sum_{n=0}^{\infty} (-1)^n
	\frac{\Gamma(-j_3+m_3+n)\Gamma(-s+n)}{\Gamma(1-s+j_1-m_1+n)} 
\, \partial^{s-n} .
\end{eqnarray}
The function $\mathcal{A}_s(z,\overline{z})$ is explicitly
given in (\ref{jaco}) 
 and it is related to the four-point function corresponding to two
highest-weight 
states with spins $j_1-\frac k2+1$ 
and  $j=\frac k2-1$ at the points $0$ and
$z$, respectively. We now show that 
monodromy 
invariant analytic continuation to non-integer values of $s$ of
equation (\ref{jaco}) 
 leads to the exact result quoted in (\ref{threepoint}).

In order to obtain the explicit expression for the three-point
function we will make use of the integral representation of the
integrand in (\ref{malda}) given in \cite{mo3}, 
namely,
\begin{eqnarray}
&& |F(a,b;c;u)|^2+\lambda |u^{1-c} F(1+b-c, 1+a-c; 2-c;u)|^2 \nonumber\\
 &&\qquad\qquad\qquad \qquad\qquad
= \frac{\gamma(c)}{\pi\gamma(b)\gamma(c-b)}|u^{1-c}|^2 \int
  d^2 t 
|t^{b-1} (u-t)^{c-b-1}(1-t)^{-a}|^2,
\label{intrh}
\end{eqnarray}
which allows to write the three-point function as
\begin{eqnarray}
\label{ser5}
 {\cal A}_3 &=& \frac{\Gamma(-s)}{\pi^2\Gamma(0)} \,
 \mathcal{S}_s(-2j_1\rho,-2j_2\rho,
\rho) \frac{\gamma(-2j_1+s)\gamma(-2j_2+s)}{\gamma(2s-2j_1-2j_2-1)} 
\frac{\Gamma(1+j_1-m_1)}{\Gamma(-j_3+m_3)}  \nonumber \\
&& \qquad\qquad\qquad
\times~
 \frac{\Gamma(1+j_1-\overline{m}_1)}{\Gamma(-j_3+\overline{m}_3)} 
\sum_{n,\overline{n}=0}^{+\infty}
 \frac{\Gamma(-j_3+m_3+n)}{\Gamma(1-s+j_1-m_1+n)} 
\frac{\Gamma(-j_3+\overline{m}_3+\overline{n})}{\Gamma(1-s+j_1-
\overline{m}_1+\overline{n})} \nonumber \\
&& \qquad\qquad\qquad \times ~
  \int d^2u \, d^2t \, u^{n} \overline{u}^{\overline{n}}
 |u|^{-2s+4j_1} 
|t|^{2s-4j_1-4j_2-4} |u-t|^{-2s+4j_2} |1-t|^{2s}.
\end{eqnarray}

Then notice that the sum in $n$ can be written in terms of yet another 
hypergeometric function as
\begin{eqnarray}
\label{ser11}
\sum_{n=0}^{+\infty} \frac{\Gamma(-j_3+m_3+n)}{\Gamma(1-s+j_1-m_1+n)} 
\, u^{n} = \frac{\Gamma(-j_3+m_3)}{\Gamma(1-s+j_1-m_1)} 
F\left(-j_3+m_3,~ 1,~ 1-s+j_1-m_1;u\right).
\end{eqnarray}
Adding the antiholomorphic dependence and completing again the
monodromy invariant 
combination we may use the integral representation (\ref{intrh}) and
write the 
three-point function in terms of the following integral:
\begin{eqnarray}
\label{ser22}
&& \int d^2u \, d^2t \, d^2z \, u^{j_1+m_1} \bar u^{j_1+\bar m_1}
(u-z)^{-1-j_2+m_2} 
(\bar u -\bar z)^{-1-j_2+\bar m_2} \nonumber \\
&& \qquad\qquad \times ~ z^{-1-j_3+m_3} \bar z^{-1-j_3+\bar m_3}
|t|^{-2-2j_{12}} 
|t-u|^{-2-2j_{13}} |1-t|^{2s}|1-z|^{-2}.
\end{eqnarray}
In order to solve this triple integral it is convenient to perform the
change of  
variables $u\rightarrow u/z$, $t\rightarrow t/z$ and integrate
$z$. Using the identity 
$F(a,b;c;t)=(1-t)^{c-a-b} F(c-a,c-b;c;t)$ and recalling that
$F(0,b;c;t)=F(a,0;c;t)=1$, 
the three-point function takes the following form:
\begin{eqnarray}
{\cal A}_3 &=& \frac {2\Gamma(-s)} {\pi^2} 
\frac{\gamma(-2j_1+s)\gamma(-2j_2+s)}{\gamma(2s-2j_1-2j_2-1)} 
\frac{\Gamma(1+j_1-m_1)}{\Gamma(-j_1+\bar m_1)} 
\frac{\Gamma(1+j_2-\bar m_2)}{\Gamma(-j_2+m_2)}
\frac{\Gamma(1+j_3-\overline{m}_3)}
{\Gamma(-j_3+\overline{m}_3)}  \nonumber \\
&& \qquad \times ~ {\cal
  S}_s(-2j_1\rho,-2j_2\rho,\rho)
  \int d^2u d^2 t~ u^{j_1+m_1} \bar u^{j_1+\bar m_1}
 (u-1)^{-1-j_2+m_2} 
(\bar u - 1)^{-1-j_2+\bar m_2}  \nonumber \\
&&~~~~~~~~~~~~~~~~~~~~~~~~~~~~
\qquad\qquad\qquad \times ~|t|^{2(-1-j_{12})} |t-u|^{2(-1-j_{13})} |1-t|^{2s}.
\label{fn}
\end{eqnarray}   

Finally this double integral can be carried out using the formula
derived in 
\cite{fukuda}, which we have collected in equation (\ref{fukuda}) in
the Appendix. 
The final result for the three-point function, after performing
Dotsenko-Fateev integral 
${\cal S}_s$ (see (\ref{dfseg}) in the Appendix) is the following:
\begin{equation}
{\cal A}_3 = -\pi^3 \,
\frac{C(-j_1,-j_2,1+j_3)W(j_1,-1-j_2,-1-j_3;m_i,\overline m_i) 
B(1+j_3)}{c^{j_1}_{m_1,\bar m_1} c^{j_2}_{m_2,\bar m_2} \prod_i(1+2j_i)},
\label{final}
\end{equation}
where the function $W(j_i, m_i,\overline m_i)$ appears in \cite{satoh} and we recall
it here for 
completeness
\begin{eqnarray}
W(j_i,m_i,\bar m_i)= \left (\frac i2\right )^2\left [ C^{12}\bar P^{12} +
C^{21}\bar P^{21} \right ],
\end{eqnarray}
with
\begin{eqnarray}
&&\left (\frac i2\right )^2 P^{12} =
  s(j_1-m_1)s(j_2-m_2)C^{31}-s(j_2-m_2)s(j_3-j_2-m_1)
C^{13}, \nonumber\\
&&C^{12}= \frac {\Gamma(-1-j_1-j_2-j_3)\Gamma(1+j_3+m_3)}{\Gamma(-j_1+m_3)}
G\left [
\begin{array}{c}
	1+j_2-m_2, ~ -j_{13}, ~ -j_3+m_3 \\ 
m_3-j_1+j_2+1, ~ -m_2-j_1-j_3
\end{array}
\right ],\nonumber\\
&& C^{31} = \frac {\Gamma(1+j_3-m_3)\Gamma(1+j_3+m_3)}{\Gamma(1+j_1+j_2+j_3)}
G
\left [
\begin{array}{c}
	1+j_2+m_2, ~ 1+j_1+j_2+j_3, ~ 1+j_1-m_1 \\ 
2-m_1+j_2+j_3, ~ 2+m_2+j_1-j_3
\end{array}
\right ],
\end{eqnarray}
and 
$C^{21}, C^{13}, P^{21}$  are obtained exchanging $j_1, m_1$ and $j_2,
m_2$ in $C^{12}, 
C^{31}$ and $P^{12}$, 
respectively. $G$ is defined by 
\begin{eqnarray}
G\left [
\begin{array}{c}
	a, ~b, ~ c \\ 
e, ~ f
\end{array}
\right ]=\frac{\Gamma(a)\Gamma(b)\Gamma(c)}{\Gamma(e)\Gamma(f)}{}_3F_2
\left [ \left .
\begin{array}{c}
	a, ~ b, ~ c \\ 
e, ~ f
\end{array}\right | 1 
\right ].
\end{eqnarray}

In order to show that our result (\ref{final}) agrees with the
expression quoted in 
\cite{satoh}, namely Eq.~(\ref{threepoint}), it is convenient to
recall the relation 
(\ref{x-m}) between $V_{j,m,\bar m}$ and $\Phi_{j,m,\bar m}$ and use
the following 
identity derived in \cite{satoh}
\begin{eqnarray}
c_{m_3,\bar m_3}^{j_3}W(j_i; m_i,\overline m_i) C(1+j_i) =B(j_3)
W(j_1, j_2, -1-j_3; m_i,\overline m_i)C(1+j_1,1+j_2,-j_3)
.
\end{eqnarray}
Similarly as in \cite{satoh} the agreement is up to a phase.

This completes the proof that the spectral flow conserving 
three-point function (\ref{final}) computed using the Coulomb gas formalism is
indeed the integral transform to the $m-$basis of the expression
computed in 
\cite{tesch1}.

\section{Spectral flow non-conserving three-point functions} 

In this section we compute the spectral flow non-conserving
 three-point function in the Coulomb gas approach. We first implement
 the
 prescription introduced in \cite{zamo, mo3} and then we propose a
 more natural novel procedure.

The method discussed in \cite{zamo} for computing spectral flow
 non-conserving $n-$point functions
starts with the evaluation of a correlator in the $x$-basis involving
  $n$ physical
 states in the $w=0$ sector and one spectral flow operator, namely,
 $\Phi_{\frac k2-1}$, 
for every unit of winding number to be violated \footnote{Notice that, up to
 a $k$-dependent factor, we can freely take $\Phi_{\frac k2-1}$ as the
 spectral flow operator instead of $\Phi_{-\frac k2}$, which was the choice
 made in \cite{mo3}, since this is its $j \leftrightarrow (-1-j)$
 reflected vertex.}. 
This correlation function is transformed to the $m$-basis assuming
 that for the spectral 
flow operator $m$ equals $\pm \frac k2$, depending on the sign of 
 winding number it 
transfers. Fourier integrals involving spectral flow operators usually
 diverge for 
these values of $m$. This problem is overcome choosing the normalization
$V_{conf}^{-1} \Phi_{\frac k2-1}$, where $V_{conf}$ 
is the volume of the
 conformal group of $S^2$ with two points fixed \cite{mo3}. 
The spectral flow operation is then
 completed dividing the correlation function by 
\begin{equation}
\prod_{i<j} (\zeta_i-\zeta_j)^{\pm \frac k2} \prod_{i,a} (\zeta_i-z_a)^{\pm m_a},
\end{equation}
where  $\zeta_i$ and  $z_a$ are the points in which the spectral
flow operators and the physical vertices are inserted, respectively, and
again the signs of the exponents depend on the sign of the $m$ of each
of the spectral 
flow operators that were introduced. After this operation, the
dependence of the result 
on the $\zeta_i$
 should disappear and the final expression is identified with a
 winding number non-conserving 
correlation function.

Among the properties of the operator $\Phi_{\frac k2-1}$, three are of
particular relevance for deriving the previous procedure.
Let us briefly recall them here. First,
$\Phi_{\frac k2-1}$ is a 
degenerate field with null descendants $J^\pm_{-1} \Phi_{\frac k2-1}$
(according to the sign of $m$) and
so every 
correlator containing it verifies an easy solvable
KZ equation \cite{zamo}. Second, 
$\Phi_{\frac k2-1}$ satisfies the following fusion rule
$\Phi_{\frac k2-1}\Phi_{j} \sim \Phi_{\frac k2-2-j}$.
Finally, the conformal dimension of $\Phi_{\frac k2-1}$ is $-\frac k4$. Then, if
this prescription is implemented in the $m-$basis, the parafermionic content of
$\Phi_{\frac k2-1}$ with $m=\pm \frac k2$ has vanishing dimension and so
it can be identified with the identity. These last two properties
allow  to interpret $\Phi_{\frac k2-1}$ as a spectral
flow operator transferring one unit of winding number.

This prescription was used in \cite{mo3} in order to explicitly obtain 
the one unit spectral flow three-point
 function.
Now we will reproduce this computation using free fields.
In order to implement this procedure in the
 Coulomb gas approach we need to translate all computations from the
 $x$-basis to the $m$-basis. So, we need to evaluate
 the coefficient of the term which goes like  $\zeta^{ m_1}
\overline{\zeta}^{\overline{m}_1}$ when $\zeta,
\overline\zeta\rightarrow 0$ in the following four-point function:
\begin{eqnarray}
\label{4}
{\cal A}_4 = \frac{\Gamma(-s)}{V_{conf}} \left\langle
\tilde{{V}}_{j_1,m_1,
\overline{m}_1}(0) \tilde{{V}}_{\frac k2-1,\frac k2,\frac k2}(\zeta) 
\tilde{{V}}_{j_2,m_2,\overline{m}_2}(1)
\tilde{{V}}_{j_3,m_3,\overline{m}_3}
(+\infty)\mathcal{S}^s\right\rangle,
\label{fpt}
\end{eqnarray}
where the conservation laws obtained from the zero mode integration of the fields are
\begin{eqnarray}
\label{5}
s=\frac k2+j_1+j_2+j_3,
\end{eqnarray}
\begin{eqnarray}
\frac k2 + m_1+m_2+m_3=0,
\end{eqnarray}
\begin{eqnarray}
\frac k2 + \overline{m}_1+\overline{m}_2+\overline{m}_3=0,
\end{eqnarray}
and all the vertex operators  create $w=0$ states.

Performing all the contractions in (\ref{fpt}) we obtain
\begin{eqnarray}
\label{casi}
&& {\cal A}_4 = \frac{\Gamma(-s)}{V_{conf}} \, |\zeta|^{-2j_1}
|\zeta-1|^{-2j_2} 
\int [dy] \prod_{i=1}^{s}
|y_i|^{-4j_1\rho}|1-y_i|^{-4j_2\rho} |\zeta-y_i|^{2}
\nonumber \\
&& ~~~~~~~~~~~~~~~~~~ \times \prod_{i<j}^s |y_i-y_j|^{4\rho} 
{\cal Q}^{-1} 
\, \partial_1 \cdots \partial_s {\cal Q} \times
\overline{\cal Q}^{-1} \, 
\overline{\partial}_1 \cdots \overline{\partial}_s \overline{\cal Q},
\end{eqnarray}
where now the $u-v$ contribution is determined by
\begin{eqnarray}
\label{qu}
{\cal Q} = \det \left[(\zeta-y_i) (1-y_i)^{-(j_2-m_2)} y_i^{j-1-j_1+m_1}\right],
\end{eqnarray}
i.e., we have ${\cal Q} = \Lambda {\cal P}$, with
\begin{eqnarray}
\Lambda = \prod_{i=1}^s(\zeta-y_i).
\end{eqnarray}

In order to compute (\ref{casi}), we first discuss the evaluation of
the ghost contribution and then perform the multiple integrations. We
will consider in the next subsection the case in which one of the
vertices, say $V_{j_1,m_1,\overline{m}_1}$, creates a
highest-weight state, i.e., we will assume that
$m_1=\overline{m}_1=j_1$ and then we will relax this assumption.

\subsection{Evaluation of the ghost correlator}

The derivatives of $\cal Q$ appearing in (\ref{casi}) when
$m_1=\overline{m}_1=j_1$ 
can be written as
\begin{eqnarray}
 \partial_1 \cdots \partial_s {\cal Q} = \det \left\{ \partial_i 
\left[(\zeta-y_i) (1-y_i)^{-j_2+m_2}
  y_i^{j-1} \right]\right\} 
= y_i^{-1} (1-y_i)^{-j_2+m_2-1} 
\sum_{l=0}^2 \ell_l^j y_i^{j-1+l},
\end{eqnarray}
where
\begin{eqnarray}
\ell_0^j= (j-1)\zeta ,
\end{eqnarray}
\begin{eqnarray}
\ell_1^j= -j + (1-j+j_2-m_2)\zeta,
\end{eqnarray}
\begin{eqnarray}
\ell_2^j= (j-j_2+m_2).
\end{eqnarray}

Therefore,
\begin{eqnarray}
\label{refere3}
{\cal Q}^{-1} \, \partial_1 \cdots \partial_s {\cal Q} = \left[
  \prod_{i=1}^s 
y_i^{-1}(\zeta-y_i)^{-1} (1-y_i)^{-1}\right] \frac{\det\left[
  \sum_{l=0}^2 \ell_l^j 
y_i^{j-1+l} \right]} {\det\left(y_i^{j-1} \right)}.
\end{eqnarray}
We can rewrite the quotient of determinants in (\ref{refere3}) as
\begin{eqnarray}
\label{pero}
\frac{\det\left[ \sum_{l=0}^2 \ell_l^j y_i^{j-1+l}
    \right]}{\det\left(y_i^{j-1} \right)} 
= \sum_{\mu} \left[ \prod_{j=1}^s \ell_{\mu_{s+1-j}}^j 
\right] \frac{\det(y_i^{j-1+\mu_{s+1-j}})}{\det\left(y_i^{j-1} \right)},
\end{eqnarray}
which formally seems to be a linear combination of Schur polynomials except
that $\mu$ is not in general a partition but a $s$-uple with entries
$0$, $1$ and $2$. However, using that the permutation of two columns
changes the sign of a determinant, it can be shown that only
partitions contribute to this sum. Indeed $s$-uples of the form
$(...,0,1,...)$ or $(...,1,2,...)$ are forbidden, and similarly those
of the form $(...,0,2,2,...)$ or $(...,0,0,2,...)$ cannot
occur. $s$-uples ending with $0$ are also forbidden. Only $s$-uples of
the form $\mu=(2,\dots,2,1,\dots,1,0,2,1,\dots,1,0,2,1,\dots,1)$
contribute and they add up to
the Schur polynomial $s_{\lambda}(y_1,\dots,y_s)$ associated to the
partition 
$\lambda = (2,\dots,2,1,\dots,1)$ up to a $(-1)^t$ factor, where $t$
is the number 
of times a couple $''0,2''$ was replaced by $''1,1''$ in $\mu$ in
order to get 
$\lambda$.

Therefore, denoting the number of $2$'s in these partitions by $n$ we can write
\begin{eqnarray}
\label{pero3}
\frac{\det\left[ \sum_{l=0}^2 \ell_l^j y_i^{j-1+l} \right]}
     {\det\left(y_i^{j-1} 
\right)} = \left[ \prod_{i=1}^s y_i \right] \sum_{n=0}^s C_{n}
     \alpha^s_{n}
(y_1,\dots,y_s),
\end{eqnarray}
where the $C_n$ depend on $\zeta$ and we have used that
\begin{equation}
s_{\lambda}(y_1,\dots,y_s) = \left[ \prod_{i=1}^s y_i \right] \times
\alpha^s_{n}
(y_1,\dots,y_s).
\end{equation}
The 
contribution of the $u-v$ fields can be written as
\begin{eqnarray}
\label{pero4}
{\cal Q}^{-1} \, \partial_1 \cdots \partial_s {\cal Q} =
\left[\prod_{i=1}^s 
(\zeta-y_i)^{-1} (1-y_i)^{-1}\right]\sum_{n=0}^s C_{n} \alpha^s_{n}(y_1,\dots,y_s).
\end{eqnarray}

It can be shown by induction on $n$ that $C_{n}$ is given by
\begin{eqnarray}
C_n = \frac{\Gamma(1+s-j_2+m_2)}{\Gamma(1+s-n-j_2+m_2)} \, \det A,
\end{eqnarray}
where
\begin{equation}
A=\left( 
\begin{array}{ccccccc}
	\ell_1^{1} & \ell_0^{2} & 0 & 0 & 0 & \cdots & 0 \\
	\ell_2^{1} & \ell_1^{2} & \ell_0^{3} & 0 & 0 & \cdots & 0 \\
	0 & \ell_2^{2} & \ell_1^{3} & \ell_0^{4} & 0 & \cdots & 0 \\
	\vdots & \vdots & \ddots & \ddots & \ddots & \vdots & \vdots \\
	0 & \cdots & 0 & 0 & \ell_2^{s-n-2} & \ell_1^{s-n-1} & \ell_0^{s-n} \\
	0 & \cdots & 0 & 0 & 0 & \ell_2^{s-n-1} & \ell_1^{s-n} \\
\end{array}
 \right).
\end{equation}

Let us denote the determinant of the submatrix formed by the first $r$
rows and $r$ 
columns of $A$ by $d_r$. Notice that det$A=d_{s-n}$. $d_r$ 
is a polynomial in $\zeta$ of degree $r$ and it is determined by the following recursion:
\begin{equation}
\label{recursion}
d_r=\ell_1^r d_{r-1} - \ell_2^{r-1} \ell_0^r d_{r-2},
\end{equation}
where $d_1=\ell_1^1$ and $d_2 = \ell_1^1 \ell_1^2 - \ell_2^1 \ell_0^2 $.
Solving (\ref{recursion}) we get
\begin{equation}
d_r= \frac{\Gamma(0)}{\Gamma(- r)} \, (1-\zeta)^{j_2-m_2},
\end{equation}
and therefore,
\begin{equation}
C_n = \frac{\Gamma(1+s-j_2+m_2)}{\Gamma(1+s-n-j_2+m_2)} \,
\frac{\Gamma(0)}
{\Gamma(-s+n)} \, (1-\zeta)^{j_2-m_2}.
\end{equation}
Note that the factor $(1-\zeta)^{j_2-m_2}$ is exactly what one needs
in order to convert the factor $(\zeta - 1)^{-j_2}$ in (\ref{casi}) into
$(\zeta -1)^{-m_2}$, as expected from
the prescription in \cite{zamo}. Let us stress here that it was
necessary to evaluate the $u-v$ contribution in order to get the
$\zeta-$dependence, unlike in \cite{zamo}, where this dependence is
deduced by looking at the $SL(2,\mathbb R)/U(1)$ coset model.

Putting everything together, the ghost contribution amounts to
\begin{eqnarray}
\label{pero42}
 {\cal Q}^{-1} \, \partial_1 \cdots \partial_s {\cal Q} &=&
\left[\prod_{i=1}^s 
(\zeta-y_i)^{-1} (1-y_i)^{-1}\right] \Gamma(0) (1-\zeta)^{j_2-m_2} \nonumber \\
&& ~~~~~~~\qquad\qquad \times ~\sum_{n=0}^s
\frac{\Gamma(1+s-j_2+m_2)}{\Gamma(1+s-n-j_2+m_2)
\Gamma(-s+n)} \, \alpha^s_{n}(y_1,\dots,y_s),
\end{eqnarray}
times a similar contribution from the antiholomorphic part. Recall
that this expression 
holds only for $m_1=\overline m_1=j_1$.

Notice that the factors $(\zeta - y_i)^{-1}$ exactly cancel their
inverse ones appearing in (\ref{casi}) and this allows us to write
the four-point function  as an Aomoto integral.

\subsection{Evaluation of the three-point function}

Replacing the result (\ref{pero42}) for the ghost correlator into (\ref{casi}) we get
\begin{eqnarray}
\label{11}
 {\cal A}_4 &=& \frac{\Gamma(-s)}{V_{conf}} \, 
\zeta^{ -j_1}
\overline{\zeta}^{ -j_1} (\zeta -1)^{-m_2} (\overline{\zeta}
-1)^
{-\overline{m}_2} 
\, \frac{\Gamma(0)}{\Gamma(-j_3+m_3)} \,
\frac{\Gamma(0)}
{\Gamma(-j_3+\overline{m}_3)}   \nonumber\\ 
&& ~~~~~ \times ~\sum_{n,\overline{n}}^s (-1)^{n+\overline{n}} \,
\frac{\Gamma(-j_3+m_3+n)}{\Gamma(-s+n)}
\frac{\Gamma(-j_3+\overline{m}_3+\overline{n})}
{\Gamma(-s+\overline{n})}\mathcal
{A}^{n,\overline{n}}_s(-2j_1\rho+1,-2j_2\rho,\rho),
\end{eqnarray}
where we have used the conservation laws (\ref{5}). Notice that the
dependence on $\zeta$ and $\overline{\zeta}$ is exactly as expected
from the  prescription in \cite{zamo}. Therefore we get the following
result for the
four-point function
involving one highest-weight state and one spectral flow operator:
\begin{eqnarray}
\label{11bisbis}
 {\cal A}_4 &=& \frac{1}{\pi} \,
 \frac{\Gamma(-s)}{\Gamma(-j_3+m_3)} \, 
\frac{\Gamma(0)}{\Gamma(-j_3+\overline{m}_3)} 
\zeta^{ -j_1}
\overline{\zeta}^{ -j_1} (\zeta -1)^{-m_2} (\overline{\zeta}
-1)^
{-\overline{m}_2} 
\nonumber\\ 
&& ~~~~~ ~~~ \times
 ~\sum_{n,\overline{n}}^s 
(-1)^{n+\overline{n}} \, \frac{\Gamma(-j_3+m_3+n)}{\Gamma(-s+n)} 
\frac{\Gamma(-j_3+\overline{m}_3+\overline{n})}
{\Gamma(-s+\overline{n})}\mathcal
{A}^{n,\overline{n}}_s(-2j_1\rho+1,-2j_2\rho,\rho).
\end{eqnarray}
The factors $\Gamma(n-s)\Gamma(\overline n-s)$ in the denominator above
allow to extended the sums to $\infty$.
Replacing the expression for ${\cal A}_s^{n,\overline
  n}(-2j_1\rho+1,-2j_2\rho,\rho)$ that
we have already found in terms of Dotsenko-Fateev integral we get, up
  to a $k$-dependent 
factor,
\begin{eqnarray}
 {\cal A}_4= \frac{\Gamma(0)}{\pi \Gamma(-s)} \, 
{\mathcal{S}}_s(-2j_1\rho+1,-2j_2\rho,\rho)\left [\zeta^{ -j_1}
(\zeta -1)^{-m_2} 
F(-j_3+m_3,\frac k2+j_1-j_2-j_3-1;-2j_3|1) 
\times \mbox{c.c.}\right ] .\nonumber
\end{eqnarray}

Using
\begin{eqnarray}
\label{15}
F(-j_3+m_3,\frac k2+j_1-j_2-j_3-1;-2j_3|1) =
\frac{\Gamma(-2j_3)\Gamma(1-\frac k2-j_1+j_2-m_3)}
{\Gamma(1-\frac k2-j_1+j_2-j_3)\Gamma(-j_3-m_3)},
\end{eqnarray}
we find 
\begin{eqnarray}
\label{19}
{\cal A}_4 &=& \frac{\Gamma(0)}{\pi \Gamma(-s)} \, 
\frac{\mathcal{S}_s(-2j_1\rho+1,-2j_2\rho,\rho)}{\gamma(1+2j_3) 
\gamma(1-\frac k2-j_1+j_2-j_3)} \nonumber\\
&&\qquad\qquad\qquad \frac{\Gamma(1+j_2+m_2)}{\Gamma(-j_2-\overline{m}_2)} 
\, \frac{\Gamma(1+j_3+\overline{m}_3)}{\Gamma(-j_3-m_3)}
\zeta^{ -j_1}
\overline{\zeta}^{ -j_1} (\zeta -1)^{-m_2} (\overline{\zeta}
-1)^
{-\overline{m}_2} .
\end{eqnarray}

Rewriting the Selberg integral in terms of $\tilde C (-j_i)$ we get
\begin{eqnarray}
\label{19bis}
{\cal A}_4 &=& \frac{\tilde{C}(-j_i)}{\gamma(-s)} \,
\prod_{i=1}^3 
\frac{\gamma((1+2j_i)\rho)}{\gamma(1+2j_i)} \,
\frac{\Gamma(1+2j_1)}{\Gamma(-2j_1)} \nonumber\\ 
&&\qquad\qquad\qquad \frac{\Gamma(1+j_2+m_2)}{\Gamma(-j_2-\overline{m}_2)} \, 
\frac{\Gamma(1+j_3+\overline{m}_3)}{\Gamma(-j_3-m_3)}\zeta^{ -j_1}
\overline{\zeta}^{ -j_1} (\zeta -1)^{-m_2} (\overline{\zeta}
-1)^
{-\overline{m}_2} .
\end{eqnarray}

Finally, using the following identity:
\begin{eqnarray}
\label{19tris}
\frac{\tilde{C}(-j_i)}{\gamma(-j_1-j_2-j_3-\frac k2)}
=\frac{1}{\rho^3}
\frac{\tilde{C}(1+j_i)}
{\gamma(j_1+j_2+j_3+3-\frac k2)} \,
\prod_{i=1}^3 
\frac{\gamma(1+2j_i)}{\gamma((1+2j_i)\rho)} , 
\end{eqnarray}
we obtain, up to $k$-dependent factors,
\begin{eqnarray}
\label{19bisbis}
{\cal A}_4 &=& \frac{\tilde{C}(1+j_i)}{\gamma(j_1+j_2+j_3+3-\frac k2)}
\, 
\frac{\Gamma(1+2j_1)}{\Gamma(-2j_1)} \nonumber\\
&&\qquad\qquad\qquad \frac{\Gamma(1+j_2+m_2)}
{\Gamma(-j_2-\overline{m}_2)} \,
\frac{\Gamma(1+j_3+\overline{m}_3)}{\Gamma(-j_3-m_3)}
\zeta^{ -j_1}
\overline{\zeta}^{ -j_1} (\zeta -1)^{-m_2} (\overline{\zeta}
-1)^
{-\overline{m}_2},
\end{eqnarray}
in complete agreement with the result in \cite{mo3}.

Recall that the last expression holds only if the state with spin
$j_1$ is a 
highest-weight state. In order to relax this condition we use the
Campbell-Backer-Hausdorff identity, namely
\begin{eqnarray}
\label{33}
 e^{\alpha J^-_0} \, V_{j,m,\overline{m}}(z) \,
e^{-\alpha J^-_0} 
= e^{\alpha [J^-_0, ~-~ ]} \, V_{j,m,\overline{m}}(z) =
\sum_{n=0}^{\infty} 
\frac{(-\alpha)^n}{n!}
\frac{\Gamma(1+j+m)}{\Gamma(1+j+m-n)} V_{j,m-n,\overline{m}}(z).
\end{eqnarray}
A similar expression is obtained if the antiholomorphic
component of the 
vertices is considered.

From the cyclic property of the trace and the commutation relation of
the screening 
operators with $J_0^-$  we get
\begin{eqnarray}
 &&\left\langle e^{\alpha J^-_0} \, V_{j_1,j_1,j_1}(0)
V_{\frac k2-1,\frac k2,\frac k2}(\zeta) \, 
e^{-\alpha J^-_0} \, V_{j_2,m_2,\overline{m}_2}(1)
V_{j_3,m_3,\overline{m}_3}(+\infty) 
\mathcal{S}^s \right\rangle  \nonumber \\
&& \qquad\qquad = \left\langle V_{j_1,j_1,j_1}(0) V_{\frac k2-1,\frac
  k2,\frac k2}(\zeta)
\, 
e^{-\alpha J^-_0} \, V_{j_2,m_2,\overline{m}_2}(1) \, e^{\alpha J^-_0}
\, 
e^{-\alpha J^-_0} \, V_{j_3,m_3,\overline{m}_4}(+\infty) \, e^{\alpha J^-_0}  
\mathcal{S}^s \right\rangle . \nonumber
\end{eqnarray}

Moreover, using the fact that $J_0^-$ annihilates $V_{-\frac k2,\frac
 k2,\frac k2}(\zeta)$ 
 we obtain
\begin{eqnarray}
\label{34bisbis}
&& \left\langle e^{\alpha J^-_0} \, V_{j_1,j_1,j_1}(0)\, e^{-\alpha
  J^-_0} 
V_{\frac k2-1,\frac k2,\frac k2}(\zeta) V_{j_2,m_2,\overline{m}_2}(1) 
V_{j_3,m_3,\overline{m}_3}(+\infty) \mathcal{S}^s \right\rangle  \nonumber \\
&& \qquad\qquad = \left\langle V_{j_1,j_1,j_1}(0) V_{\frac k2-1,\frac k2,\frac k2}(\zeta)
  \, 
e^{-\alpha J^-_0} \, V_{j_2,m_2,\overline{m}_2}(1) \, e^{\alpha J^-_0}
  \, 
e^{-\alpha J^-_0} \, V_{j_3,m_3,\overline{m}_3}(+\infty) \, e^{\alpha J^-_0}  
\mathcal{S}^s \right\rangle .\nonumber
\end{eqnarray}

So the four point function satisfies the following identity:
\begin{eqnarray}
\label{37}
&& \sum_{{n_1}=0}^{\infty} \frac{(-\alpha)^{n_1}}{n_1!}
\frac{\Gamma(1+2j_1)}
{\Gamma(1+2j_1-n_1)} \left\langle V_{j_1,j_1-n_1,j_1}(0)
V_{\frac k2-1,\frac k2,\frac k2}(\zeta) 
V_{j_2,m_2,\overline{m}_2}(1) V_{j_3,m_3,\overline{m}_3}(+\infty)
\mathcal{S}^s 
\right\rangle \nonumber \\
&& ~~~~~ \qquad = \sum_{{n_2,n_3}=0}^{\infty} \frac{\alpha^{n_2+n_3}}{n_2!
  n_3!} 
\frac{\Gamma(1+j_2+m_2)}{\Gamma(1+j_2+m_2-n_2)}
\frac{\Gamma(1+j_3+m_3)}
{\Gamma(1+j_3+m_3-n_3)} \nonumber \\
&& ~~~~~\qquad \qquad \qquad \times ~\left\langle V_{j_1,j_1,j_1}(0)
V_{\frac k2-1,\frac k2,\frac k2}(\zeta) 
V_{j_2,m_2-n_2,\overline{m}_2}(1)
V_{j_3,m_3-n_3,\overline{m}_3}(+\infty) \mathcal{S}^s 
\right\rangle.
\end{eqnarray}

Identifying powers on both sides of (\ref{37}) we find
\begin{eqnarray}
\label{39}
&& \left\langle V_{j_1,j_1-n_1,j_1}(0) V_{\frac k2-1,\frac k2,\frac k2}(\zeta)
V_{j_2,m_2,
\overline{m}_2}(1) V_{j_3,m_3,\overline{m}_3}(+\infty) \mathcal{S}^s
\right\rangle 
\nonumber \\
&& \qquad\qquad\qquad = \sum_{p=0}^{n_1} (-1)^{n_1} \left( 
\begin{array}{c}
	n_1 \\ p
\end{array}
\right) \,
\frac{\Gamma(1+2j_1-n_1)}{\Gamma(1+2j_1)} \, 
\frac{\Gamma(1+j_2+m_2)}{\Gamma(1+j_2+m_2-p)} \,
\frac{\Gamma(1+j_3+m_3)}
{\Gamma(1+j_3+m_3-n_1+p)} \nonumber \\
&& \qquad \qquad\qquad ~~~~~~~~~~~\times ~
\left\langle V_{j_1,j_1,j_1}(0) 
V_{\frac k2-1,\frac k2,\frac k2}(\zeta) 
V_{j_2,m_2-p,\overline{m}_2}(1) 
V_{j_3,m_3-n_1+p,\overline{m}_3}(+\infty) \mathcal{S}^s
\right\rangle ,
\end{eqnarray}
i.e.,
\begin{eqnarray}
\label{39trique}
&& \left\langle V_{j_1,m_1,j_1}(0) V_{\frac k2-1,\frac k2,\frac k2}(\zeta)
V_{j_2,m_2,
\overline{m}_2}(1) V_{j_3,m_3,\overline{m}_3}(+\infty) \mathcal{S}^s
\right\rangle 
\nonumber \\
&& \qquad~~ = \sum_{p=0}^{\infty} (-1)^{j_1-m_1} 
\frac{\Gamma(1+j_1+m_1)\Gamma(1+j_3+m_3)}{p!\Gamma(1+2j_1)\Gamma(-j_1+m_1)
\Gamma(-j_2-m_2)} \,
\frac{\Gamma(-j_2-m_2+p) \Gamma(-j_1+m_1+p)}
{\Gamma(1+j_3+m_3-j_1+m_1+p)} \nonumber \\
&& \qquad ~~~~~~~~~~~~~~~~~ \times ~
\left\langle V_{j_1,j_1,j_1}(0) 
V_{\frac k2-1,\frac k2,\frac k2}(\zeta) 
V_{j_2,m_2-p,\overline{m}_2}(1) 
V_{j_3,m_3-j_1+m_1+p,\overline{m}_3}(+\infty) \mathcal{S}^s
\right\rangle .
\end{eqnarray}

Taking into account the antiholomorphic component
we obtain
\begin{eqnarray}
\label{39final}
&& \left\langle V_{j_1,m_1,\overline{m}_1}(0) V_{\frac k2-1,\frac
  k2,\frac k2}(\zeta)
V_{j_2,m_2,\overline{m}_2}(1) V_{j_3,m_3,\overline{m}_3}(+\infty)
\mathcal{S}^s 
\right\rangle  \nonumber \\
&&  = \sum_{p,\overline{p}=0}^{\infty} (-1)^{j_1-m_1} (-1)^{j_1-\overline{m}_1}
\frac{\Gamma(1+j_1+m_1)\Gamma(1+j_3+m_3)}{p!\Gamma(1+2j_1)\Gamma(-j_1+m_1)
\Gamma(-j_2-m_2)} \,
\frac{\Gamma(-j_2-m_2+p) \Gamma(-j_1+m_1+p)}
{\Gamma(1+j_3+m_3-j_1+m_1+p)} \nonumber \\
&& \qquad \qquad\times ~
\frac{\Gamma(1+j_1+\overline{m}_1)\Gamma(1+j_3+\overline{m}_3)}{\overline{p}!
\Gamma(1+2j_1)\Gamma(-j_1+\overline{m}_1)\Gamma(-j_2-\overline{m}_2)} \,
\frac{\Gamma(-j_2-\overline{m}_2+\overline{p}) \Gamma(-j_1+\overline{m}_1+\overline{p})}
{\Gamma(1+j_3+\overline{m}_3-j_1+\overline{m}_1+\overline{p})}  \nonumber \\
&& \qquad \qquad\times ~
\left\langle V_{j_1,j_1,j_1}(0) 
V_{\frac k2-1,\frac k2,\frac k2}(\zeta) 
V_{j_2,m_2-p,\overline{m}_2-\overline{p}}(1) 
V_{j_3,m_3-j_1+m_1+p,\overline{m}_3-j_1+\overline{m}_1+\overline{p}}(+\infty) 
\mathcal{S}^s
\right\rangle .
\end{eqnarray}

It follows that
\begin{eqnarray}
\label{39finalbis}
 {\cal A}_4 &=& \frac{\tilde{C}(1+j_i)}{\gamma(j_1+j_2+j_3+3-\frac k2)} \,
\sum_{p,\overline{p}=0}^{\infty}
\frac{\Gamma(1+j_1+m_1)\Gamma(1+j_3+m_3)}{p!~\Gamma(-j_1+m_1)} \,
 \Gamma(-j_1+m_1+p)\Gamma(1+j_2+m_2)
\nonumber \\
&& \qquad \times ~ 
\frac{\Gamma(1+j_1+\overline{m}_1)\Gamma(1+j_3+\overline{m}_3)}{\overline{p}!~
\Gamma(-j_1+\overline{m}_1)\Gamma(-j_2-\overline{m}_2)} \,
 \Gamma(-j_1+\overline{m}_1+\overline{p})
\zeta^{-j_1} \overline{\zeta}^{-j_1} (\zeta-1)^{-m_2+p} 
(\overline{\zeta}-1)^{-\overline{m}_2+\overline{p}}  \nonumber \\
&=& \frac{\tilde{C}(1+j_i)}{\gamma(j_1+j_2+j_3+3-\frac k2)}
\frac{\Gamma(1+j_2+m_2)}
{\Gamma(-j_2-\overline{m}_2)} \frac{
  \Gamma(1+j_1+m_1)\Gamma(1+j_3+\overline{m}_3)}
{\Gamma(-j_1-\overline{m}_1)\Gamma(-j_3-m_3)} 
\zeta^{-m_1} 
(\zeta-1)^{-m_2}\times c.c. \nonumber\\
\end{eqnarray}
where we have used that, for any $a \in \mathbb R$,
\begin{equation}
F(-n, a, a|z) = (1-z)^n .
\end{equation}
Therefore
\begin{eqnarray}
\label{ulti}
&& {\cal A}_3^{w=1} =
\frac{\tilde{C}(1+j_i)}{\gamma(j_1+j_2+j_3+3-\frac k2)} 
\frac{
  \Gamma(1+j_1+m_1)}{\Gamma(-j_1-\overline{m}_1)}
\frac{\Gamma(1+j_2+m_2)}{\Gamma(-j_2-\overline{m}_2)} 
\frac{
\Gamma(1+j_3+\overline{m}_3)}
{\Gamma(-j_3-m_3)},
\end{eqnarray}
where, following the prescription in \cite{mo3, zamo}, we have cancelled
the $\zeta$ dependence in the four point function in order to get the
one unit spectral flow three-point function 

This result,
worked out in the free field approach,
agrees with the exact expression reported in \cite{mo3} (with the
obvious convention
changes, see (\ref{3pointwind})). 
It has been obtained for one state in the discrete series.
 However it also holds for the continuous series. This may
be seen by noting that the three-point function (\ref{ulti}) does not change upon
reflection of the vertex operator according to the quantum
version of (\ref{asymp}) \footnote{This requires replacing the factor
  $1+2j$ in (\ref{asymp}) by ${\cal R}(j)=[\rho(1-2j)\gamma(\rho(1-2j))]^{-1}
$.}. Although the dependence of (\ref{ulti}) on
$s$ is not obvious, it is exclusively contained in the factor
$\gamma(s-k+3)$ in the denominator. Therefore, the analytic
continuation to non-integer $s$ presents no difficulties. To our
knowledge, this is the first independent verification of the exact
result obtained in \cite{mo3} for the one unit
spectral flow three-point function for generic states.

\subsection{Alternative method to compute spectral flow non-conserving amplitudes}

In the previous section we computed a four-point
function involving only $w=0$ states. It is straightforward to see
that the result still holds for states in arbitrary winding sectors as
long as
$\sum_i w_i=0$.  Even though this condition is consistent with the prescription in
\cite{mo3, zamo}, the natural way to introduce winding violation in
the Coulomb gas approach is through a modification of the conservation laws.
In this section we argue for an alternative procedure to compute
winding 
number non-conserving correlators.

Recall that  non-vanishing $n-$point functions must verify the following constraints
\cite{becker}
\begin{eqnarray}
\sum_{i=1}^N(j_i-m_i-w_i)-s&=&-1,\label{lcu}\\
\sum_{i=1}^N(j_i-m_i)-s&=&-1,\label{lcv}\\
\sum_{i=1}^N(j_i+\frac{k-2}2w_i)-s&=&-
1 .\label{cl}
\end{eqnarray} 

It is possible to achieve $\sum_iw_i= \pm 1$ inserting
 the following auxiliary operators 
\begin{eqnarray}
 \eta^- (\zeta)& = & e^{iv(\zeta)},\\
\eta^+ (\zeta)& = & e^{(k-2)u(\zeta)}e^{-i(k-1)v(\zeta)}e^{\sqrt{2(k-2)}\phi(\zeta)}. 
\end{eqnarray}
These  vertices correspond to  operators with 
quantum numbers $j=\frac k2-1, m=\pm\frac k2,
 w=\mp 1$
(see (\ref{vcw})). $\eta^\mp$ can be
 thought of as identities in the
 $w=\mp 1$ sectors. This is emphasised by the fact that they have zero
 conformal dimension.
 Even though they have non-trivial OPE with physical
 vertex operators, namely 
\begin{eqnarray}
V^{w}_{j,m,\bar m}(z) e^{iv(\zeta)} & \sim & (z-\zeta)^{j-m}
V_{j+\frac k2 -1, m+\frac k2, w-1}(\zeta) ,\\
V^{w}_{j,m,\bar m}(z) e^{(k-2)u(\zeta)}e^{-i(k-1)v(\zeta)}
e^{\sqrt{2(k-2)}\phi(\zeta)} & \sim & (z-\zeta)^{-j-m}
V_{j+\frac k2 -1, m-\frac k2, w+1}(\zeta),
\end{eqnarray}
the physical states on the left- and right-hand sides of these
expressions can be identified using the equivalence between ${\cal D}^{\pm,
  w\mp 1}_j$ and ${\cal D}^{\mp, w }_{\frac k2-2-j}$. 
Moreover,
the operators $\eta^\pm$ commute with the currents up to a singular
 state.
Notice the
similarity of  these properties  and those of the spectral
flow operator proposed in \cite{zamo}.

Since physical correlators should not depend upon the points where
these operators are inserted, we can freely assume that they are
located at infinity, consistently with a modification of the
background charge in (\ref{lcu})$-$(\ref{cl}).

We now show that this prescription reproduces the expected result for the
one unit spectral flow three-point function.
Let us start by computing the four-point function
\begin{eqnarray}
&&{\cal A}_4=\langle V_{j_1 m_1 \bar m_1 }^{w_1}(0)
V_{j_2 m_2 \bar m_2 }^{w_2}(1)V_{j_3 m_3 \bar m_3}^{w_3}(\infty)\eta^-(\zeta)
{\cal S}^s\rangle ,\label{3ptsw}
\end{eqnarray} 
which leads to the following  conservation laws 
\begin{eqnarray}
&&s=\sum_i j_i+\frac k2 ,\\
 &&\sum_i m_i + \frac k2=0 ,\\
 && \sum_i w_i = 1 .
\end{eqnarray}

The contribution from $\phi$ is not modified with respect to the
spectral flow
conserving case, i.e., it gives (\ref{fi}). The $u-v$ correlator now
contributes
\begin{eqnarray}
&&\zeta^{(-j_1+m_1)}(\zeta-1)^{(-j_2+m_2)}\overline 
\zeta^{(-j_1+\overline m_1)}(\overline \zeta-1)^{(-j_2+\overline
  m_2)}\nonumber\\
&&~~~~~~~~~~~~~~~~~~~\prod_{i=1}^s|y_i|^{-2w_1}|1-y_i|^{-2w_2}
{\cal P}^{-1}\partial _1\cdots \partial_s {\cal Q}~
\overline {\cal P}^{-1}\partial _1\cdots \partial_s \overline
{\cal Q}
\label{nues}
\end{eqnarray}
where ${\cal P}$ and ${\cal Q}$ are given by (\ref{pe}) and (\ref{qu}), respectively.

Putting everything together, it is easy to see that the dependence of the four-point
function on
$\zeta$ exactly cancels. Indeed,
the quotient of
determinants  in (\ref{nues}) can be read from (\ref{pero42}) 
in the case $m_1=\overline m_1=j_1$.
In order to relax the highest-weight
condition all the steps performed in the previous section go through
to this case up to equation (\ref{39final}). Therefore  the
one unit spectral flow three-point function is obtained directly in
this way, through well defined operators.

So far we have seen that the operator $\eta^-$
 reproduces the prescription presented in \cite{zamo} and used
in \cite{mo3}, 
in the particular case of one winding unit. Now we will show that
this statement 
is general.

Let us assume that we are computing a correlator in the sector $w=N$. 
We need to evaluate expectation values of the form \footnote{Of course
this three-point function vanishes for $|N|> 1$. Here we are
interested in comparing the $\zeta-$dependence with the prescription in \cite{zamo}}
\begin{eqnarray}
\mathcal{A}^{w=N}_3 = \Gamma(-s) \left\langle
\tilde{\mathcal{V}}^{w_1}_{j_1,m_1,
\overline{m}_1}(0) \tilde{\mathcal{V}}^{w_2}_{j_2,m_2,\overline{m}_2}(1) 
\tilde{\mathcal{V}}^{w_3}_{j_3,m_3,\overline{m}_3}(+\infty)
\prod_{i=1}^{N_+} \eta^+(\zeta^+_i)
\prod_{l=1}^{N_-} \eta^-(\zeta^-_{l})
\mathcal{S}^s\right\rangle ,
\end{eqnarray}
where $w=\sum_iw_i=N=N_--N_+$.

The contribution from contractions of $\phi$ is the following
\begin{eqnarray}
\prod_{l=1}^{N_+} |\zeta_l^+|^{-4j_1} |1-\zeta_l^+|^{-4j_2}
\prod_{r<l} 
|\zeta_r^+ - \zeta_l^+|^{-4(k-2)} \prod_{i=1}^s |y_i|^{-4j_1\rho} 
|1-y_i|^{-4j_2\rho} \prod_{l=1}^{N_+} |\zeta_l^+ - y_i|^{4} 
\prod_{i<j}^s|y_i-y_j|^{4\rho},
\end{eqnarray}
whereas
contractions of the $u-v$ fields give
\begin{eqnarray}
&&\prod_{m=1}^{N_-} (\zeta^-_m)^{-(j_1-m_1)}
  (1-\zeta^-_m)^{-(j_2-m_2)} \prod_{m<r} 
(\zeta_m^- - \zeta_r^-)  \prod_{l=1}^{N_+}
  (\zeta^+_l)^{(j_1-m_1)} 
(1-\zeta^+_l)^{(j_2-m_2)} \nonumber \\
&& ~~~~~~~ \times \prod_{l<n} (\zeta_l^+ - \zeta_n^+)^{2k-3} \times
  \prod_{l=1}^{N_+} 
\prod_{m=1}^{N_-} (\zeta_l^+ - \zeta^-_m)^{-(k-1)} \times
  {\mathcal{P}}^{-1} ~ 
\partial_1\dots \partial_s (\Lambda \mathcal{P}) 
\end{eqnarray}
times the antiholomorphic contribution, where
\begin{eqnarray}
\mathcal{P} = \det \left[ \prod_{l=1}^{N_+} (\zeta_l^+-y_i)^{-k+2}
  (1-y_i)^{-j_2+m_2} 
y_i^{j-1-j_1+m_1} \right]
\end{eqnarray}
and now
\begin{eqnarray}
\Lambda = \prod_{i=1}^s \prod_{m=1}^{N_-} (\zeta_m^--y_i)
\prod_{l=1}^{N_+} 
(\zeta_l^+-y_i)^{-1}.
\end{eqnarray}
We have omitted the powers involving $w_i$ since they cancel.

Putting everything together we get
\begin{eqnarray}
\mathcal{A}^{w=N}_3 &=& \Gamma(-s) \prod_{m=1}^{N_-}
  (\zeta^-_m)^{-j_1+m_1} 
(\overline{\zeta}^-_m)^{-j_1+\overline{m}_1} (1-\zeta^-_m)^{-j_2+m_2} 
(1-\overline{\zeta}^-_m)^{-j_2+\overline{m}_2}  \prod_{m<r} (\zeta_m^-
  - \zeta_r^-)
(\overline{\zeta}_m^- - \overline{\zeta}_r^-)  \nonumber \\
&& ~~ \times ~\prod_{l=1}^{N_+} (\zeta^+_l)^{-j_1-m_1} 
(\overline{\zeta}^+_l)^{-j_1-\overline{m}_1} (1-\zeta^+_l)^{-j_2-m_2} 
(1-\overline{\zeta}^+_l)^{-j_2-\overline{m}_2}  \prod_{l<n} 
(\zeta_l^+ - \zeta_n^+)(\overline{\zeta}_l^+ - \overline{\zeta}_n^+)  \nonumber \\
&& ~~ \times ~\prod_{m=1}^{N_-} \prod_{l=1}^{N_+} (\zeta_l^+ -
  \zeta^-_m)^{-(k-1)} 
(\overline{\zeta}_l^+ - \overline{\zeta}^-_m)^{-(k-1)} \int
       [dy] \prod_{i=1}^s 
|y_i|^{-4j_1\rho} |1-y_i|^{-4j_2\rho} \prod_{l=1}^{N_-} |\zeta_l^- -
y_i|^{4} 
 \nonumber \\
&& ~~~~~~~~~~~~ \times ~ \prod_{i<j}^s|y_i-y_j|^{4\rho}
 {\mathcal{P}}^{-1} ~ 
\partial_1\dots \partial_s (\Lambda \mathcal{P}) 
\times   \overline {\mathcal{P}}^{-1} ~ 
\overline\partial_1\dots \overline\partial_s (\overline\Lambda \overline{\mathcal{P}}) ,
\end{eqnarray}
which can be recast as
\begin{eqnarray}
 \mathcal{A}^{w=N}_3 & = & \prod_{m=1}^{N_-} (\zeta^-_m)^{m_1} 
(\overline{\zeta}^-_m)^{\overline{m}_1} (1-\zeta^-_m)^{m_2} 
(1-\overline{\zeta}^-_m)^{\overline{m}_2}  \prod_{m<r} 
(\zeta_m^- - \zeta_r^-)^{k/2}(\overline{\zeta}_m^- -
  \overline{\zeta}_r^-)^{k/2} 
 \nonumber \\
&& ~~~~~~~\times ~\prod_{l=1}^{N_+} (\zeta^+_l)^{-m_1}
 (\overline{\zeta}^+_l)^{-\overline{m}_1} 
(1-\zeta^+_l)^{-m_2} (1-\overline{\zeta}^+_l)^{-\overline{m}_2}
 \prod_{l<n} 
(\zeta_l^+ - \zeta_n^+)^{k/2} (\overline{\zeta}_l^+ -
 \overline{\zeta}_n^+)^{k/2}  
 \nonumber \\
&& ~~~~~~~ \times ~\prod_{l=1}^{N_+} \prod_{m=1}^{N_-} (\zeta_l^+ -
 \zeta^-_m)^{-k/2} 
(\overline{\zeta}_l^+ - \overline{\zeta}^-_m)^{-k/2} ~\times
 ~\mathcal{A}^{w=0}_{N+3} ,
\label{a3wn}
\end{eqnarray}
with
\begin{eqnarray}
\mathcal{A}^{w=0}_{N+3} &=& \Gamma(-s) \prod_{m=1}^{N_-}
|\zeta^-_m|^{-2j_1} 
|1-\zeta^-_m|^{-2j_2} \prod_{m<r} |\zeta_m^- - \zeta_r^-|^{-(k-2)} 
 \prod_{l=1}^{N_+} |\zeta^+_l|^{-2j_1} |1-\zeta^+_l|^{-2j_2}
\nonumber \\
&& \times~  \prod_{l<n} 
(\zeta_l^+ - \zeta_n^+)^{-(k-2)}  
\prod_{m=1}^{N_-} \prod_{l=1}^{N_+} |\zeta_l^+ -
 \zeta^-_m|^{-(k-2)} 
 \nonumber\\
&&\times \int [dy] \prod_{i=1}^s |y_i|^{-4j_1\rho} |1-y_i|^{-4j_2\rho}
\prod_{l=1}^{N_+} 
|\zeta_l^+ - y_i|^{4} \prod_{i<j}^s|y_i-y_j|^{4\rho}\left [
\frac{1}{\mathcal{P}} ~ 
\partial_1\dots \partial_s (\Lambda \mathcal{P}) \times
\mbox{c.c.}\right ] . \nonumber\\
\end{eqnarray}

We can now identify $\mathcal{A}^{w=0}_{N+3}$ with a $(N+3)-$point
 function involving three physical states with winding number adding
 up to zero and $N$ spectral flow operators, namely
\begin{eqnarray}
 \mathcal{A}^{w=0}_{N+3} &=& \Gamma(-s) \prod_{m=1}^{N_-}
  |\zeta^-_m|^{-2j_1} 
|1-\zeta^-_m|^{-2j_2} \prod_{m<r} |\zeta_m^- - \zeta_r^-|^{-(k-2)}
\prod_{l=1}^{N_+} |\zeta^+_l|^{-2j_1} |1-\zeta^+_l|^{-2j_2}
\nonumber \\
&& \times ~\prod_{l<n} 
|\zeta_l^+ - \zeta_n^+|^{-(k-2)}  
 \prod_{m=1}^{N_-} \prod_{l=1}^{N_+} |\zeta_l^+ -
\zeta^-_m|^{-(k-2)} 
 \int [dy] \prod_{i=1}^s |y_i|^{-4j_1\rho} |1-y_i|^{-4j_2\rho} \nonumber \\
&& \times ~\prod_{m=1}^{N_-} |\zeta_m^+ - y_i|^{2} \prod_{l=1}^{N_+} 
|\zeta_l^+ - y_i|^{2} \prod_{i<j}^s|y_i-y_j|^{4\rho} \left [
\frac{1}{\mathcal{Q}} ~ 
\partial_1\dots \partial_s \mathcal{Q} \times \mbox{c.c.}\right ],
\end{eqnarray}
where
\begin{eqnarray}
\mathcal{Q} = \Lambda \mathcal{P} = \det \left[\prod_{m=1}^{N_-}
  (\zeta_m^--y_i) 
\prod_{l=1}^{N_+} (\zeta_l^+-y_i)^{-(k-1)} (1-y_i)^{-j_2+m_2} y_i^{j-1-j_1+m_1}\right].
\end{eqnarray}

We thus see that the operators $\eta^\pm$ produce the effect
 described in the previous section, since the $\zeta^\pm$ factors
 in ${\cal A}_3^{w=N}$ in (\ref{a3wn}) are precisely those that are to be
 cancelled ad-hoc according to the prescription in \cite{zamo}.
This shows that our approach completely agrees
with the one 
suggested in \cite{zamo}, with the additional advantage that here it is
implemented 
through well defined operators.

\section{Summary and conclusions}

We have proved that a proper treatment of
 the background charge method reproduces
the generic three-point functions in the $SL(2,\mathbb{R})$ WZNW model. 
Indeed,  we have found complete agreement with all
previous exact results obtained from the bootstrap approach \cite{mo3,
 tesch1, zamo, tesch2}. Our work completes previous calculations
 performed in 
\cite{becker, gn2}
where the Coulomb gas formalism was used to compute spectral flow
 conserving three-point functions containing
at least one state of the discrete highest-weight series or two
 highest-weight states in
 the spectral flow non-conserving case, respectively. 
Indeed,
the highest-weight condition considered in \cite{becker, gn2} allowed to simplify
the computation of the $\beta -\gamma$ 
contribution to the three-point functions
 and it also permitted the use of the well known Dotsenko-Fateev integrals.
However, 
 while the analytic continuation 
to global descendant discrete states performed in
\cite{becker} gives results in accordance with those of \cite{mo3, tesch1},
as shown in \cite{satoh}, the more general case had not been considered
before.

Here we have been able to compute
the ghost contribution
for generic states. We showed that it can be expressed in terms of Schur
polynomials and we solved the resulting new integrals of Dotsenko-
Fateev type, namely we computed Aomoto integrals in the complex plane.
Finally, we used monodromy invariance
to properly perform the analytic continuation to non-integer
number of screening operators. Actually, we proved that all spectral
flow conserving
 three-point functions can be obtained acting with a suitable
 differential operator
 on a four-point-like function and thus showed that a monodromy
 invariant compatible
analytic continuation is required. The techniques  developed in Section
 3 to deal
 with the spectral flow conserving three-point function were then applied
 to solve the one unit spectral flow three-point function in Section 4. In all
 cases we
 verified agreement with known exact results. Moreover, we proposed a novel method
to compute winding number non-conserving $n-$point functions through
well defined operators and we
showed that it reproduces the prescription suggested in \cite{zamo},
with the advantage that ad-hoc cancellations are not required.
In fact, applying this method to the case $n=3$ we have shown that the result
is independent of the insertion point of the auxiliary operator, which suggests
that it is possible to compute the
 one unit spectral flow three-point function
in the free field approach through a modification of the background
charge.

Having realized that the ghost contribution to the
amplitudes can be expressed in terms of
Schur polynomials and the resolution of Aomoto
integrals in the complex plane are interesting byproducts of our work.
These are important ingredients 
 of other related problems in CFT and they will certainly be
elements of the Coulomb gas computation of higher-point functions, 
though  in a more complex version. 
Actually one important simplification in the three-point function
is the appearance of a minimal partition. Instead, 
Schur polynomials do not reduce to the elementary symmetric polynomial
in the case of an extra insertion point, as we have seen in the
computation of the four-point function in section 4. This implies that 
the multiple integrals appearing in the four-point
function not only get more involved because of the extra insertion
point  but also because they include a more
complicated Schur polynomial. 
Nevertheless, even though it might seem unlikely to obtain 
explicit expressions for generic four-point functions, given that they
are not available in the simpler Liouville theory,
our results provide a step forward towards the resolution
of the factorization of  AdS$_3$ four-point functions and
the determination of unitarity and full consistency of the theory
using Coulomb gas techniques.

Indeed,
beyond the formal aspects regarding the
 explicit confirmation of the validity of the free field 
method 
 and the eventual mathematical justification of the Coulomb
gas formalism
in this non-RCFT, other important applications of our work
are related to string theory on $AdS_3$. Even if closed expressions
for higher than three-point functions cannot be found,
we expect that the techniques  we
have developed will enable us to analyse
the consistency of this theory  studying the short
distance limit of four-point functions. It might be argued that
the background charge
method appears unnecessarily complicated to achieve this task, 
given that the spacetime Lie
and conformal
symmetries  allowed to resolve the problem exactly in the related $H_3^+$ model. 
In fact, factorization and crossing symmetry 
have been proved in \cite{tesch1, tesch2} using the relation between KZ and
BPZ equations. Moreover, the analytic continuation of these results to string
theory on AdS$_3$ was performed in \cite{mo3}, where the factorization
of the four-point function of $w=0$ states was shown to be consistent
both with the Hilbert space proposed in \cite{mo1} and with the
winding violation pattern arising from the theory of $SL(2, \mathbb
R)$ representations. However, the inclusion of non-trivial spectral
flow sectors simplifies in the formalism that we have presented here. 
In fact, whereas only one unit
spectral flow vertex operators have been constructed in the $x-$basis,
it is easy to include winding number in the $m-$basis. 
We believe that all these arguments
justify the validation of
the Coulomb gas approach up to the known exact correlators and the
development of the new computational techniques which we have achieved in this
article.

Finally, the
extension of our results both to higher-point functions and to the
supersymmetric case, started in \cite{hn}, might be relevant to the
scope of verifying the $AdS_3$/CFT$_2$ correspondence, 
another relevant issue which has not been completed yet
(see \cite{gaberdiel, pakman, ps} for some recent work in this direction).

\bigskip

\noindent{\bf Acknowledgments}: C.N. is grateful to V. Fateev and G. Giribet 
for correspondence and to P. Minces for reading the manuscript.
This work was supported by PROSUL under contract CNPq
490134/2006-8, CONICET, Universidad de Buenos Aires and ANPCyT.

\bigskip

\section{\bf Appendix: useful formulas}

\medskip
In this appendix we collect some of the formulas we have used in our 
computations. 

\medskip

\noindent{\bf 1. Selberg integral and Dotsenko-Fateev formula.}

\medskip

The following integral was first derived by Selberg in \cite{selberg}:

\begin{eqnarray}
S_s(a,b,\rho) & = & \int_0^1 dy_1 \cdots \int_0^1 dy_s 
\prod_{i=1}^s y_i^{a-1} (1-y_i)^{b-1} \prod_{i<j} 
|y_i-y_j|^{2\rho} \nonumber \\
& = & \prod_{i=0}^{s-1} \frac{\Gamma(a+i\rho) \Gamma(b+i\rho) 
\Gamma((i+1)\rho+1)}{\Gamma(a+b+(s+i-1)\rho) \Gamma(\rho+1)}.
\end{eqnarray}

The extension of  Selberg integral to the complex plane 
was carried out by Dotsenko and Fateev in \cite{df2}. 
They obtained the following result:

\begin{eqnarray}
\mathcal{S}_s(a,b,\rho) & = & \int \prod_{i=1}^n d^2
y_i \prod_{i=1}^s |y_i|^{2a-2} |1-y_i|^{2b-2}
\prod_{i<j} |y_i-y_j|^{4\rho}  \nonumber \\
& = & s! \pi^s \gamma(1-\rho)^s \prod_{i=1}^s \gamma(i\rho) 
\prod_{i=0}^{s-1} \gamma(a+i\rho) \gamma(b+i\rho) \gamma(1-a-b-(s-1+i)\rho).
\label{dfseg}
\end{eqnarray}

\medskip

\noindent{\bf 2. Aomoto integrals of order $k$.}

\medskip
In \cite{aomoto} Aomoto computed a family of integrals 
generalizing Selberg's. Aomoto's integral of order $k$ 
is defined as
\begin{eqnarray}
A^k_s(a,b,\rho) & = & \int_0^1 dy_1 \cdots \int_0^1 dy_s 
\alpha_k^s(y_1,\dots,y_s)  
\prod_{i=1}^s y_i^{a-1} (1-y_i)^{b-1} 
\prod_{i<j} |y_i-y_j|^{2\rho} ,
\end{eqnarray}
where $\alpha_k^s(y_1,\dots,y_s)$ is the elementary symmetric 
polynomial of order $k$, i.e.,
\begin{eqnarray}
\alpha_k^n(y_1,\dots,y_n)  =  \sum_{1\le j_1< \cdots < j_k \le n} \, 
\prod_{1=1}^k y_{j_i} 
=  \frac{1}{k! (n-k)!} \sum_{\sigma_n} \prod_{i=1}^k y_{\sigma_n(i)},
\end{eqnarray}
and the last sum is made over the permutations of order $n$.

The following result was obtained in \cite{aomoto}:
\begin{eqnarray}
\label{hhh}
A^k_s(a,b,\rho) & = & \left( 
\begin{array}{c}
	s \\ k
\end{array}
\right) \frac{\Gamma(\alpha+s) \Gamma(\alpha+\beta+2s-k-1)}{\Gamma(\alpha+s-k) 
\Gamma(\alpha+\beta+2s-1)} S_s(a,b,\rho),
\end{eqnarray}
where $\alpha=a/\rho$ and $\beta=b/\rho$.

Aomoto integrals can be conveniently arranged in a single 
expression. Let us consider the following integral:
\begin{eqnarray}
\label{ec4}
A_s(a,b,\rho;z) & = & \int_0^1 dy_1 \cdots \int_0^1 dy_s 
\prod_{i=1}^s y_i^{a-1} (1-y_i)^{b-1} (z-y_i) \prod_{i<j} |y_i-y_j|^{2\rho}
.
\end{eqnarray}

Notice that $A_s(a,b,\rho;z)$ is an $s$-degree polynomial in the variable $z$.
Using Newton identities, namely,
\begin{eqnarray}
\prod_{i=1}^s (z-y_i)  =  \sum_{k=0}^s (-1)^k 
\alpha^s_k(y_1,\dots,y_s) z^{s-k} 
 =  \sum_{k=0}^s (-1)^{s-k} \alpha^s_{s-k}(y_1,\dots,y_s) z^{k} ,
\end{eqnarray}
it is easy to see that
\begin{eqnarray}
A_s(a,b,\rho;z) & = & \sum_{k=0}^s (-1)^k A^k_s(a,b,\rho) z^{s-k} ,
\end{eqnarray}
i.e., Aomoto integral of order $k$ is, up to a phase, 
the coefficient of the $(s-k)$-degree term of $A_s(a,b,\rho;z)$.
This can be more conveniently written in terms of monic Jacobi
polynomials as
\begin{eqnarray}
\label{ec5678}
A_n(a,b,\rho;z) & = & \frac{(-1)^n}{2^n} S_n(a,b,\rho;z) 
\overline{P}_n^{\alpha-1,\beta-1}(1-2z) ,
\end{eqnarray}
where we have used the definitions collected below.
\medskip

\noindent{\bf 3. Jacobi polynomials.}
\medskip

Recall the following
expression of Jacobi polynomials (see \cite{tabla}):
\begin{eqnarray}
P_n^{\alpha,\beta}(x) & = & \frac{\Gamma(\alpha+n+1)}{\Gamma(\alpha+1)
\Gamma(n+1)}\, {}_2F_1\left(-n,\alpha+\beta+n+1;\alpha+1;\frac{1-x}{2}\right) 
 \nonumber \\
& = & \frac{\Gamma(\alpha+n+1)}{\Gamma(n+1)\Gamma(\alpha+\beta+n+1)} 
\sum_{k=0}^n \left( 
\begin{array}{c}
	n \\ k
\end{array}
\right) \frac{\Gamma(\alpha+\beta+n+k+1)}{\Gamma(\alpha+k+1)} 
\left( \frac{x-1}{2}\right)^k  .
\end{eqnarray}

Monic Jacobi polynomials read
\begin{eqnarray}
\label{ec3}
\overline{P}_n^{\alpha,\beta}(x) & = & 2^n \frac{\Gamma(\alpha+n+1)
\Gamma(\alpha+\beta+n+1)}{\Gamma(\alpha+1)
\Gamma(\alpha+\beta+2n+1)}\, {}_2F_1\left(-n,\alpha+\beta+n+1;\alpha+1;
\frac{1-x}{2}\right)  \nonumber \\
& = & 2^n \frac{\Gamma(\alpha+n+1)}
{\Gamma(\alpha+\beta+2n+1)} \sum_{k=0}^n \left( 
\begin{array}{c}
	n \\ k
\end{array}
\right) \frac{\Gamma(\alpha+\beta+n+k+1)}{\Gamma(\alpha+k+1)} 
\left( \frac{x-1}{2}\right)^k \nonumber \\
& = & 2^n \frac{\Gamma(n+1)\Gamma(\alpha+\beta+n+1)}
{\Gamma(\alpha+\beta+2n+1)} P_n^{\alpha,\beta}(x)  .
\end{eqnarray}

\vfill\eject

\noindent{\bf 4. Fukuda-Hosomichi integral.}
\medskip

The following double integral has been derived in
\cite{fukuda}:
\begin{eqnarray}
W(\alpha_i,\bar\alpha _i,\alpha_i ', \bar\alpha_i ',\sigma) &=&
\int d^2 z d^2 w ~ z^{\alpha_1} (1-z)^{\alpha_2} \bar z^{\bar\alpha_1}
(1-\bar z)^
{\bar\alpha_2 }
w^{\alpha_1 '} (1-w)^{\alpha_2 '} \bar w^{\bar\alpha_1 '}(1-\bar w)^
{\bar\alpha_2 '}
|z-w|^{4\sigma}\nonumber\\
&=&\left ( \frac i2\right )^2\left \{ C^{12}[\alpha_i,\alpha_i ']
P^{12}[\bar\alpha_i, \bar\alpha_i '] + C^{21}[\alpha_i, \alpha_i ']
P^{21}[\bar\alpha_i,\bar\alpha_i '] \right \} , \label{fukuda}
\end{eqnarray}
where
\begin{eqnarray}
\label{53b}
 C^{ab}[\alpha_i,
\alpha_i '] = \frac {\Gamma(1+\alpha_a+\alpha '_a-k')
\Gamma(1+\alpha_b+\alpha_b '-k')}{\Gamma(k'-\alpha_c-\alpha_c ')}
G\left [
\begin{array}{c}
	\alpha_a '+1, ~ \alpha_b+1, ~ k'-\alpha_c-\alpha_c ' \\ 
1-\alpha_c + \alpha_a ', ~ \alpha_b-\alpha_c ' +1
\end{array}
\right ] ,
\end{eqnarray}
and 
\begin{eqnarray}
\left (\frac i2\right )^2\left [ \begin{array}{c}
P^{12}\\P^{21} \end{array}
\right ]=
A_\beta\left [ \begin{array}{c}
C^{23}\\C^{32}\end{array} \right ]=
A_\alpha^T\left [ \begin{array}{c}
C^{31}\\C^{13}\end{array} \right ] ,  
\end{eqnarray}
with
\begin{eqnarray}
 G\left [ \begin{array}{c}
a,b,c\\e,f \end{array}
 \right ]=\frac {\Gamma(a)\Gamma(b)\Gamma(c)}{\Gamma(e)\Gamma(f)}
{_3} F_2\left [ \left .\begin{array}{c}
a,b,c\\ e,f \end{array} \right | 1 \right ],\nonumber
\end{eqnarray}
\begin{eqnarray}
\alpha_1+\alpha_2+\alpha_3+1 &=& k'= ~-2\sigma-1,\nonumber\\
\alpha_1 '+\alpha_2 '+\alpha_3 '+1 & = & k'= ~-2\sigma-1,\nonumber
\end{eqnarray}
\begin{eqnarray}
A_\alpha \left [ 
\begin{array}{cc} 
s(\alpha)s(\alpha ')&-s(\alpha)s(\alpha '-k')\\
-s(\alpha ')s(\alpha - k')& s(\alpha)s(\alpha ')
\end{array}
\right ]  ,\nonumber
\end{eqnarray}
and $s(x)=sin(\pi x)$.

\end{document}